\documentclass[a4paper,fleqn,usenatbib]{mnras}

\usepackage{newtxtext,newtxmath,placeins,afterpage}

\usepackage[T1]{fontenc}
\usepackage{ae,aecompl}

\usepackage{graphicx,color}	
\usepackage{amsmath}	
\usepackage{caption}
\usepackage{subcaption}

\def\hi{\mbox{H\sc{i}}}

\def\atlas{{{ATLAS}}$^{\rm 3D}$}

\def\co{$^{12}$CO}
\def\tco{$^{13}$CO}
\def\ceo{C$^{18}$O}
\def\stack{\textsc{Stackarator}}

\def\kms{km\,s$^{-1}$}
\def\msun{M$_{\odot}$}

\def\arcsec{$^{\prime \prime}$}
\definecolor{Mygrey}{gray}{0.75}

\newcommand{\ltsimeq}{\raisebox{-0.6ex}{$\,\stackrel{\raisebox{-.2ex}{$\textstyle <$}}{\sim}\,$}}
\newcommand{\gtsimeq}{\raisebox{-0.6ex}{$\,\stackrel{\raisebox{-.2ex}{$\textstyle >$}}{\sim}\,$}}

\mathchardef\mhyphen="2D

\title[VERTICO: CO isotopologues ]{VERTICO IX: Signatures of environmental processing of the gas in Virgo cluster spiral galaxies through mapping of CO isotopologues} 
\author[Timothy A. Davis et al.]{\parbox{\textwidth}{\vspace{-0.6cm}
Timothy A. Davis,$^{1}$\thanks{E-mail: DavisT@cardiff.ac.uk},
Toby Brown,$^{2}$
Mar\'ia J. Jim\'enez-Donaire,$^{3,4}$
Christine D. Wilson,$^{5}$
Dhruv Bisaria,$^{6}$
Alessandro Boselli,$^{7}$
Barbara Catinella,$^{8,9}$
Aeree Chung,$^{10}$
Luca Cortese,$^{8,9}$
Sara Ellison,$^{11}$
Bumhyun Lee,$^{12,10}$
Ian D. Roberts,$^{13,14}$
Kristine Spekkens,$^{6}$
Vicente Villanueva,$^{15}$
and Nikki Zabel.$^{16}$}
\vspace{0.4cm}\\
\parbox{\textwidth}{
$^{1}$Cardiff Hub for Astrophysics Research \&\ Technology, School of Physics \&\ Astronomy, Cardiff University, Queens Buildings, Cardiff, CF24 3AA, UK\\
$^{2}$Herzberg Astronomy and Astrophysics Research Centre, National Research Council of Canada, 5071 West Saanich Rd, Victoria, BC V9E 2E7, Canada\\
$^{3}$Observatorio Astron\'omico Nacional (IGN), C/Alfonso XII, 3, 28014 Madrid, Spain\\
$^{4}$Centro de Desarrollos Tecnológicos, Observatorio de Yebes (IGN), 19141 Yebes, Guadalajara, Spain\\
$^{5}$Department of Physics and Astronomy, McMaster University, 1280 Main Street West, Hamilton, Ontario L8S 4M1, Canada\\
$^{6}$Department of Physics, Engineering Physics, and Astronomy, Queen's University, Kingston, ON K7L 3N6, Canada\\
$^{7}$Aix Marseille Univ, CNRS, CNES, LAM, Marseille, France\thanks{Scientific associate INAF - Osservatorio Astronomico di Cagliari, Via della Scienza 5, 09047 Selargius (CA), Italy}\\
$^{8}$International Centre for Radio Astronomy Research, The University of Western Australia, 35 Stirling Hwy, 6009 Crawley, WA, Australia\\
$^{9}$ARC Centre of Excellence for All Sky Astrophysics in 3 Dimensions (ASTRO 3D), Australia\\
$^{10}$Department of Astronomy, Yonsei University, 50 Yonsei-ro, Seodamun-gu, Seoul 03722, Republic of Korea\\
$^{11}$Department of Physics \& Astronomy, University of Victoria, PO Box 1700 STN CSC, Victoria, BC V8W 2Y2, Canada\\
$^{12}$Korea Astronomy and Space Science Institute, 776 Daedeokdae-ro, Daejeon 34055, Republic of Korea\\
$^{13}$Department of Physics \& Astronomy, University of Waterloo, Waterloo, ON N2L 3G1, Canada\\
$^{14}$Waterloo Centre for Astrophysics, University of Waterloo, 200 University Ave W, Waterloo, ON N2L 3G1, Canada\\
$^{15}$Departamento de Astronom{\'i}a, Universidad de Concepci{\'o}n, Barrio Universitario, Concepci{\'o}n, Chile\\
$^{16}$Department of Astronomy, University of Cape Town, Private Bag X3, Rondebosch 7701, South Africa
}}
\date{Accepted 2025 July 23. Received 2025 June 20; in original form 2025 January 23}
\pubyear{2025}
\begin{document}
\label{firstpage}
\pagerange{\pageref{firstpage}--\pageref{lastpage}}
\maketitle

\begin{abstract}
In this work we study CO isotopologue emission in the largest cluster galaxy sample to date: 48 VERTICO spiral galaxies in Virgo.  We show for the first time in a significant sample that the physical conditions within the molecular gas appear to change as a galaxy's ISM is affected by environmental processes. \tco\ is detected across the sample, both directly and via stacking, while \ceo\ is detected in a smaller number of systems. 
We use these data to study trends with global and radial galaxy properties. We show that the \co/\tco\ line ratio {changes systematically with a variety of galaxy properties, including mean gas surface density, \hi-deficiency and galaxy morphology.}  \tco/\ceo\ line ratios vary significantly, both radially and between galaxies, suggesting real variations in abundances are present. Such abundance changes may be due to star formation history differences, or speculatively even stellar initial mass function variations. We present a model where the optical depth of the molecular gas appears to change as a galaxy's ISM is affected by environmental processes. The molecular gas appears to become more transparent as the molecular medium is stripped, and then more opaque as the tightly bound remnant gas settles deep in the galaxy core. This explains the variations we see, and also helps explain similar observations in cluster early-type galaxies. 
Next generation simulations and dedicated observations of additional isotopologues could thus provide a powerful tool to help us understand the impact of environment on the ISM, and thus the quenching of galaxies. 
\end{abstract}




\section{Introduction}
\label{intro}
Galaxy clusters are extreme environments, that are known to play an active role in the ongoing evolution of their constituent galaxies. It has long been known that galaxies in clusters are redder, and more likely to have elliptical and lenticular morphologies, than those found in lower density environments \citep[e.g.][]{1974ApJ...194....1O,1980ApJ...236..351D,1984ApJ...285..426B}. 
We believe that these changes in the cluster galaxy population are due to environmental quenching effects, such as ram pressure stripping, tidal stripping and thermal evaporation  \citep[e.g.][]{1972ApJ...176....1G,1980ApJ...237..692L,1996Natur.379..613M}. Depending on the relative strength of these processes they can remove the circumgalactic medium around galaxies (leading to strangulation; \citealt{1977Natur.266..501C}), or also the colder ISM, leading to direct quenching of star formation \citep[e.g.][]{2014A&A...564A..67B,2019MNRAS.483.2251Z,2021PASA...38...35C}.

Given the above, it is clearly crucial to understand the impact that different environmental mechanisms have on the cold gas within galaxies. This allows us to distinguish which mechanisms are important in different environments, and thus how star formation is suppressed in dense cluster cores. The atomic gas within galaxies is well known to be strongly affected by environment \citep[e.g.][]{1973MNRAS.165..231D,1980A&A....83...38C,2005A&A...429..439G,2009AJ....138.1741C}. However, this material is less closely linked to star formation. The cold molecular interstellar medium (ISM) is the key component within which stars are formed, but is almost always located deep within the potential well of the galaxy, where environmental effects are more subtle \citep{1986ApJ...310..660S,2006PASP..118..517B,2008A&A...491..455V,2014A&A...564A..67B,2017MNRAS.466.1382L,2019MNRAS.483.2251Z,2022ApJS..263...40M}. 

The Virgo Environment Traced in CO \citep[VERTICO;][]{Brown2021} program, of which this paper is part, is one of the most recent efforts to understand the effects of environment on the cold molecular ISM.  VERTICO is an ALMA (Atacama Large Millimeter/submillimeter Array) Compact Array Large Program probing the molecular gas content at sub-kpc scales in a statistically significant number of disc galaxies in the Virgo cluster. 

VERTICO allows us to study the effect of the cluster environment on the cold gas in galaxies using observations of carbon monoxide (\co), the second most abundant molecule after H$_2$. Initial studies with this dataset have, for instance, revealed the effects of HI-identified environmental mechanisms on molecular gas \citep{2022ApJ...933...10Z, Watts2023}, how star formation proceeds in cluster galaxies \citep{2022arXiv221116521J}, and the gas distribution and star-formation efficiency of member galaxies \citep{2022ApJ...940..176V}.

In this work we attempt to unravel the effects of the cluster environment on the cold gas in galaxies in a different way: using the less abundant isotopologues of carbon monoxide, \tco\ and \ceo. Isotopologue studies allow us to probe the physical conditions within the gas, study the enrichment of the ISM, and can even help in deciphering the star formation history of galaxies \citep[e.g.][]{1991A&A...249...31S,1992A&A...264...55C,1999RPPh...62..143W,2008ApJ...679..481P,2014MNRAS.445.2378D,2017ApJ...836L..29J,2018MNRAS.475.3909C,2022A&A...662A..89D,2023ApJS..268....3C}. All of these properties are potentially altered in dense environments. Despite this, very few studies focused on CO isotopologues in cluster environments have been carried out. \cite{2015MNRAS.450.3874A} is one exception, which showed that \tco-to-\co\ ratios are higher in cluster early-type galaxies (ETGs) than those in the field, potentially due to their star formation histories or differences in the structure of the ISM. \cite{2018ApJ...866L..10L} also used this method to study the gas conditions within a single ram-pressure stripped galaxy. Here we can extend these works to consider the impact of the cluster on the ISM of a sample of cluster spiral galaxies.

Specifically, the Atacama Compact Array (ACA) data from VERTICO allows us to study the \co(2-1) and \ceo(2-1) transitions at sub-kpc scales within 48 spiral galaxies in the Virgo cluster. \tco(2-1) observations are available for 34 of these objects. The \co\ molecule is the most abundant, and as such is usually optically thick everywhere within galaxies. \tco\ and \ceo, on the other hand, are significantly less abundant (by factors of $\approx$70; \citealt{1993A&A...274..730H}, and $\approx$200; \citealt{1991A&A...249...31S}, respectively). This means these lines are typically optically thin (always for \ceo, and in all but the densest regions in the case of \tco), allowing us to estimate the optical depth of the \co\ emitting molecular material, and search for abundance differences which may be caused by environmental processes. These data can be compared with the EMPIRE \citep[EMIR Multiline Probe of the ISM Regulating Galaxy Evolution;][]{2019ApJ...880..127J} and CLAWS \citep[CO isotopologue Line Atlas within the Whirlpool galaxy Survey;][]{2022A&A...662A..89D} and EDGE-CALIFA (Extragalactic Database for Galaxy Evolution - Calar Alto Legacy Integral Field Area) surveys \citep{2023ApJS..268....3C}, which also studied these isotoplogues at similar resolution in nearby (mostly) field spiral galaxies, and the (resolved and unresolved) works on these isotopologues in ETGs \citep[e.g.][]{2010MNRAS.407.2261K,2012MNRAS.421.1298C,2013MNRAS.433.1659D,2015MNRAS.450.3874A,2016MNRAS.463.4121T}.

This paper is organized as follows: in Section \ref{data} we discuss the data used in this work, and our methodology, including the new stacking software we have developed. In Section \ref{results} we present our results, before discussing them in Section \ref{discuss}, and concluding in Section \ref{conclude}. Throughout we assume a distance of 16.5 Mpc for all galaxies within the Virgo cluster \citep{2007ApJ...655..144M}.
\section{Data and methodology}
\label{data}

\begin{figure}
\begin{subfigure}{0.47\textwidth}
	\includegraphics[width=\textwidth,trim=0.2cm 0cm 0cm 0cm,clip]{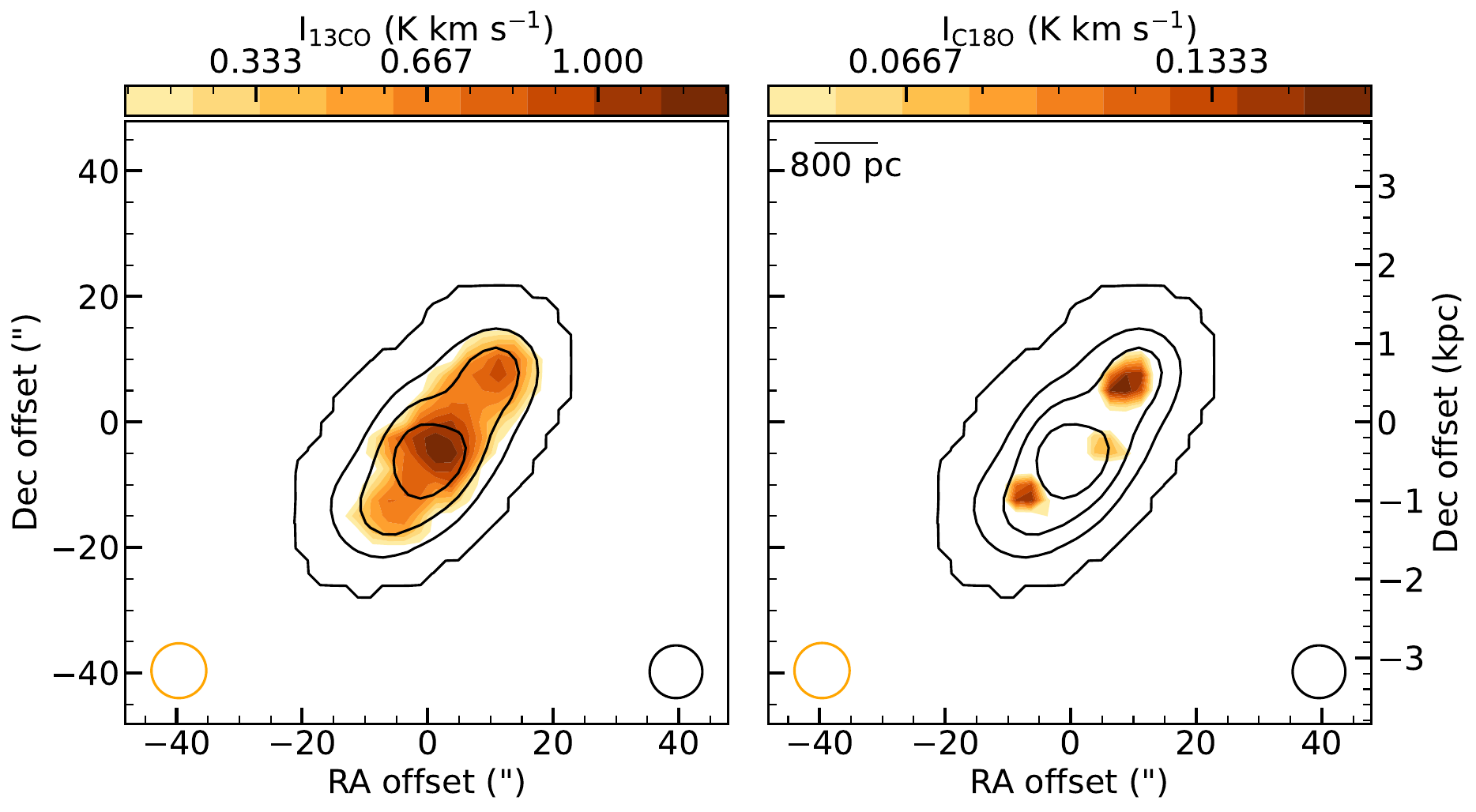}
    \caption{IC3392}
\end{subfigure} 
\begin{subfigure}{0.47\textwidth}
	\includegraphics[width=\textwidth,trim=0.2cm 0cm 0cm 0cm,clip]{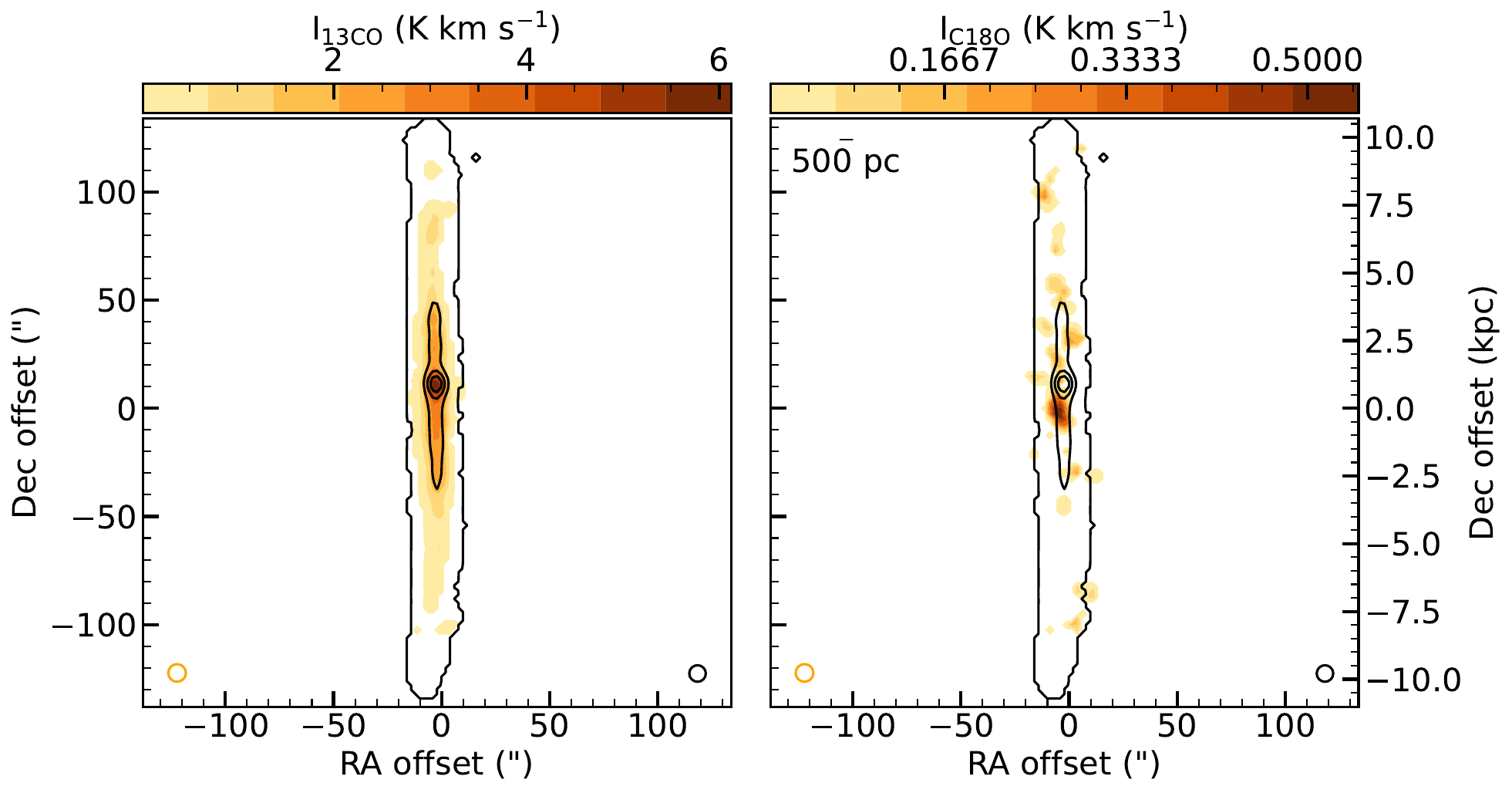}
    \caption{NGC4302}
\end{subfigure} 
\begin{subfigure}{0.47\textwidth}
	\includegraphics[width=\textwidth,trim=0.2cm 0cm 0cm 0cm,clip]{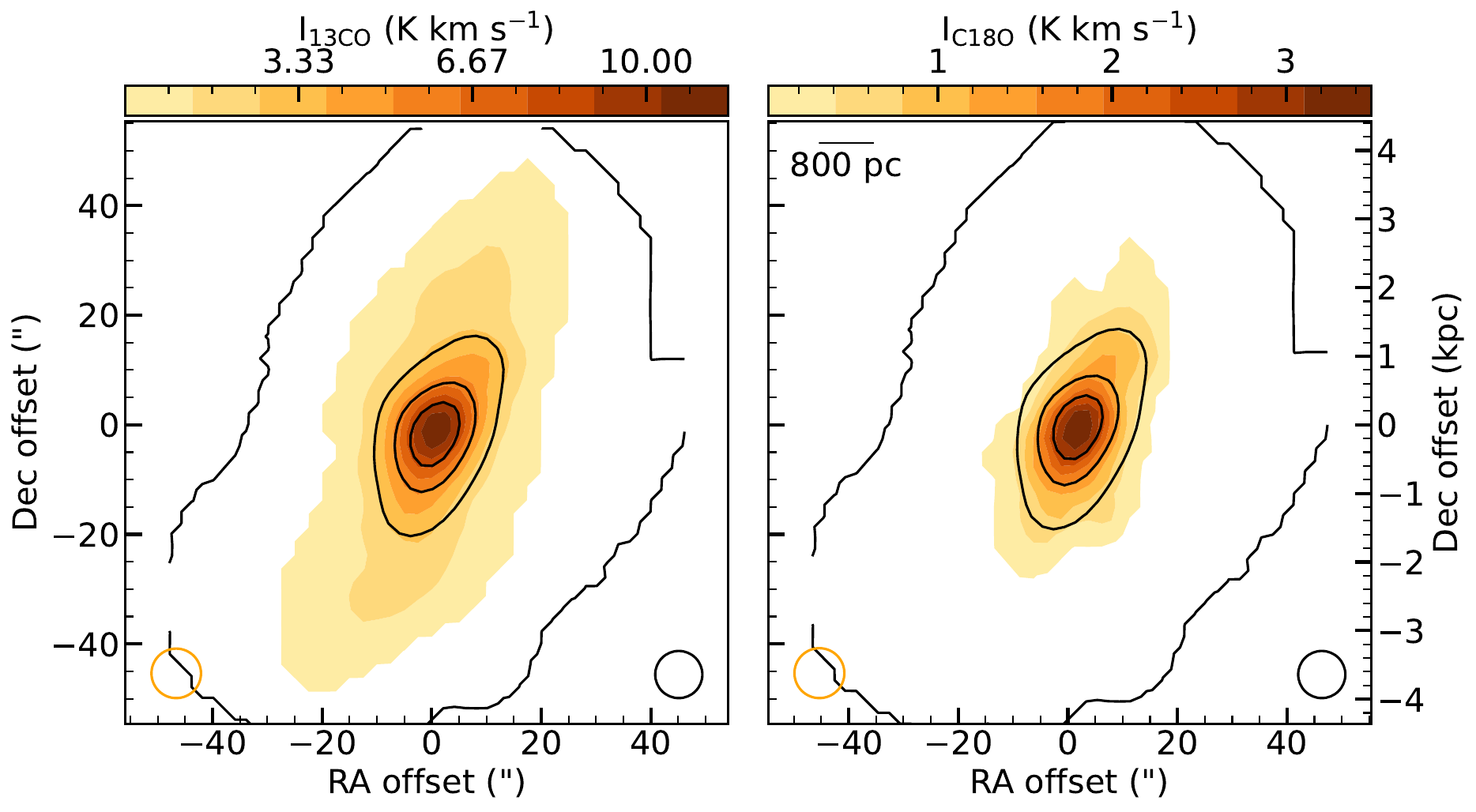}
    \caption{NGC4568}
\end{subfigure}
	\caption{\tco(2-1) and \ceo(2-1) integrated intensity (moment zero) maps for the three systems where both lines were detected without stacking. The \tco\ detection is shown in the left panel, and \ceo\ in the right panel. 
	In each case five black contours delineate the \co(2-1) emitting region, starting at a column density of 5 \msun\,pc$^{-2}$, and spaced equally to the maximum.  The isotopologue emission is shown in orange in ten contours spaced from 10\% of the peak emission to 100\%. 	
The corresponding  \co\ and isotopologue telescope beams are shown as black and orange ellipses in the bottom right and left corner of each image, respectively.  }
	\label{fig:directdetect}
\end{figure}

\subsection{Sample}
The full VERTICO\footnote{\url{https://www.verticosurvey.com/}} \citep{Brown2021} sample consists of a total of 51 Virgo Cluster galaxies, selected by the Very Large Array Imaging of Virgo in Atomic Gas survey \citep[VIVA,][]{2009AJ....138.1741C} to study how the environment may be affecting them, via ram pressure stripping, starvation, and tidal interaction. These sources cover a wide range of star formation properties and span two orders of magnitude in stellar mass (9$<$log(M$_{*}$/M$_{\odot}$)$<$11). 
We here study the 48 galaxies in the VERTICO survey that are detected in \co(2-1) and where at least one CO isotopologue was in the bandpass.

\subsection{ALMA data}

The observations for VERTICO were carried out using the ACA during Cycle 7 (2019.1.00763.L.). Out of the 51 galaxies in the sample, 36 targets were newly observed, while 15 galaxies were already surveyed and the data were publicly available in the ALMA archive \citep{Cramer2020,Leroy2021a}.  
Total Power (TP) observations were included to recover extended CO emission at large angular scales for the most extended 25 out of 36 targets. 
For each galaxy, a Nyquist-sampled mosaic was obtained in order to map their molecular gas disk. The nominal flux calibration uncertainty of ALMA in Band 6 during Cycle 7 was 5-10\% according to the ALMA Cycle 7 Technical Handbook. A detailed description of the data reduction and imaging process is available in \citet{Brown2021}.

The data observed as part of the VERTICO program included spectroscopic observations of the ALMA Band 6 continuum, and the $^{12}$CO\,(2-1) transition (rest frequency 230.542\,GHz), as well as the isotopologues $^{13}$CO\,(2-1) at a rest frequency of 220.402\,GHz and C$^{18}$O\,(2-1) at a rest frequency of 219.564\,GHz. However, the data available from the archive did not include \tco(2-1), as this was not possible in previous ALMA cycles. 
In this work our results are derived from the original native resolution cubes as delivered from the ACA, which have a typical beamsize of $\approx$7\arcsec ($\approx$560\,pc).
 
 Integrated intensity maps are computed directly from these final data products, implementing a masking process described in \cite{Sun2018}. This method employs a spatially and spectrally varying noise, which we measure in every pixel and channel before primary beam correction. For that, a mask is generated by combining a core mask, selecting spaxels with $\mathrm{S/N}\geq 3.5$ in three consecutive channels or more, and a wing mask for spaxels with $\mathrm{S/N}\geq 2$ in two consecutive channels or more.  Moment maps are then created from the original datacubes with the mask applied. We refer the reader to the VERTICO survey paper \cite{Brown2021} for a detailed description of this procedure.

\begin{figure*}
	\includegraphics[width=1\textwidth,trim=0.2cm 0cm 0cm 0cm,clip]{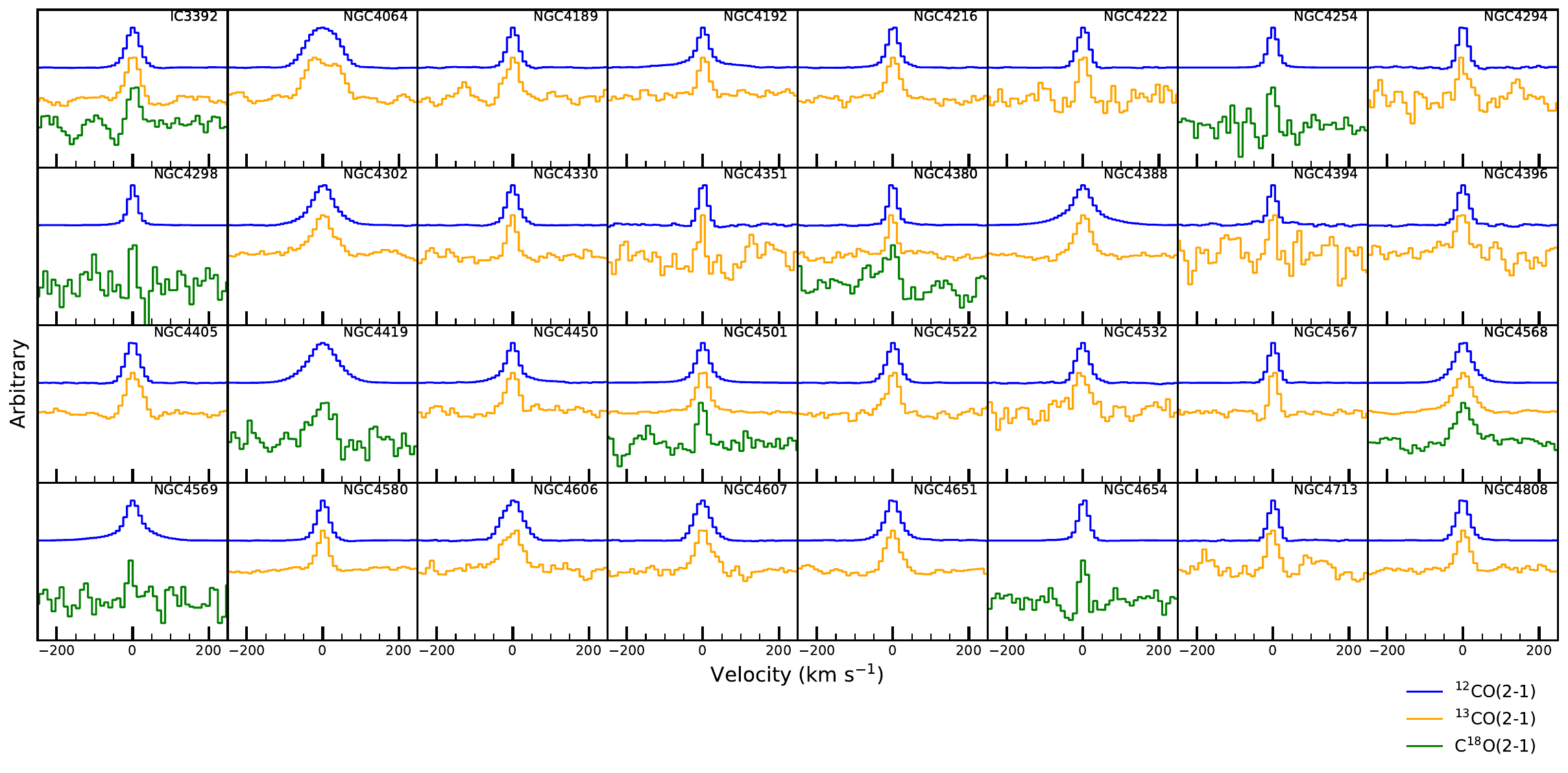}\\
	\caption{\co(2-1), \tco(2-1) and \ceo(2-1) stacked spectra  (shown in blue, orange and green, respectively) for entire VERTICO galaxies where at least two isotopologues were detected. Galaxies without \tco\ detections are from archival data where this line was not in the bandpass. We note that the \co\ spectra here are not the same as those shown in \protect \cite{Brown2021} because of the stacking procedure used (see Section \ref{stack_describe}).  }
	\label{fig:allgals_stack}
\end{figure*}

\subsection{\stack}
\label{stack_describe}
Isotopologue emission is directly detected in the integrated intensity maps of a number of our sample galaxies. However, to extend the sample further we also conduct a stacking analysis, using a newly developed code called \stack. 
This code allows image-plane stacking of spectral-line data, based upon the velocity field observed in another tracer. 
A full description of this code is included in Appendix \ref{stack_appendix}.

Here we use \stack\ in two modes. Firstly we shift the spectra from every \co(2-1) detected spaxel within each galaxy to a common velocity reference according to the \co(2-1) moment one (intensity-weighted mean velocity) maps presented in \cite{Brown2021}, before co-adding them. We note that the resulting spectrum thus loses its normal double-horned shaped, and becomes approximately gaussian, with line wings present from pixels with strong velocity gradients. These shifted spectra give a global measurement of the CO isotopologue fluxes for each system, discussed in Sections \ref{stackedglobal} to \ref{tcoandceo}. We then proceed to stack the data within concentric elliptical beam-width sized apertures in each system (again based on the \co(2-1) observed velocity maps), allowing us to study radial gradients in the isotopologues in Section \ref{sec:radprof}. 
We note that we did attempt to stack using the velocities observed in the VIVA \hi\ data, as the \hi\ is often more extended than the \co, but found that this made a negligible difference in the returned isotopologue line fluxes. See \cite{2022ApJ...940..176V} for an application of a similar tehcnique for the \co\ in our sample galaxies.

For each stack (be it a global, or from a single radial bin) we fit the spectrum to determine if a line is detected. 
The stacked \co\  (and sometimes the \tco) lines are non-gaussian in shape, due to beam smearing in the inner parts of the galaxy where the rotation curve is steep. At a given position within the galaxy beam smearing creates a skewed line profile, however when all lines of sight are combined the line profile becomes symmetric, but with large wings. The \ceo\ lines should in principle have a similar line shape, but the signal-to-noise on this line is not high enough for this to be well constrained.

We fit the lines using a two step process, fitting both a single gaussian model, and a double gaussian (where the secondary gaussian is constrained to have the same centre, but a larger velocity width). We note that Gauss Hermite functions would provide another possible way to fit these profiles, but at the signal to noise reached here a gaussian approach is sufficient.
These models are fitted to the data using Markov chain Monte Carlo (MCMC) fitting code \textsc{GAStimator}\footnote{https://github.com/TimothyADavis/GAStimator}.
The best fits returned are then compared by computing the Bayesian information criterion (BIC), which naturally penalises more complex models. The double gaussian is only assumed where the BIC indicates this is preferable (i.e. BIC$_{\rm double}$<BIC$_{\rm single}$), in all other cases the single gaussian model is used.  The total model fluxes are then estimated by integrating the best model. We tested if more stringent criteria (e.g. the difference in the BIC values needing to be $>10$) improved the fits, but this did not significantly change our results.

Given the correlated non-uniform noise present in stacked spectra it is important that the uncertanties are correctly estimated. We do this using a Monte-Carlo "look-elsewhere" technique. We sum the stacked spectrum at 1000 randomly chosen off-line locations using a window the same width as the best fit gaussian full-width at zero intensity. The standard deviation of the resulting integrated intensities at these random centroid velocities can then be used to define the uncertanty in the integrated intensity measurement, and thus how significantly this emission is detected.

\section{Results}
\label{results}

\subsection{Direct detections}
We directly detect both \tco\ and \ceo\ emission in the moment zero maps for three of our VERTICO galaxies (Figure \ref{fig:directdetect}). Each plot has the \tco\ detection in the left panel, and \ceo\ in the right panel. The isotopologue emission is shown as ten orange filled contours, starting from the 5$\sigma$ level and extending to the brightest integrated intensity peak, as shown in the colour bars. On each panel we overplot the \co\ contours as a guide to the eye. 

We detect and map \tco\ directly in another 14 systems where \ceo\ is not detected (see Appendix Figure \ref{fig:directdetect_appendix} and \ref{fig:directdetect_appendix2}). NGC4501 has especially bright and well resolved \tco\ emission, as shown in Figure \ref{fig:directdetect_appendix}, panel (h). The spatial distributions of these isotopologues will be discussed further in Section \ref{discuss_directdetect}.

\subsection{Stacked global detections}
\label{stackedglobal}
To further enhance our sample size we stack our VERTICO data-cubes for each system as described in Section \ref{stack_describe}. The resulting stacked emission lines for galaxies with at least one isotopologue detected are shown in Figure \ref{fig:allgals_stack}. Each panel of this figure shows a single galaxy, with the stacked \co, \tco\ and \ceo\ emission lines in blue, orange and green respectively. Spectra are arbitrarily scaled on the y-axis for visibility. Where a \tco\ line is not shown this is because the archival data did not cover this frequency range. For \ceo, on the other hand, lines not shown are not detected at 5$\sigma$.  We do not detect either isotopologue in seven systems, which are thus not shown in this figure. We detect \tco\ in our global stack for every other galaxy where data was available. \ceo, on the other hand, was detected in 9 galaxies. The \co, \tco\ and \ceo\ integrated intensities of each system retrieved through our stacking analysis are listed in Table \ref{Table1}. We note that the stacked line NGC4064 is especially broad, and the \tco\ detection is double-peaked. The cause of this is unclear, but may relate to the bar, or known ionised gas outflow \citep{2006AJ....131..747C}.

\begin{figure}
	\includegraphics[width=0.485\textwidth,trim=0.2cm 0cm 0cm 0cm,clip]{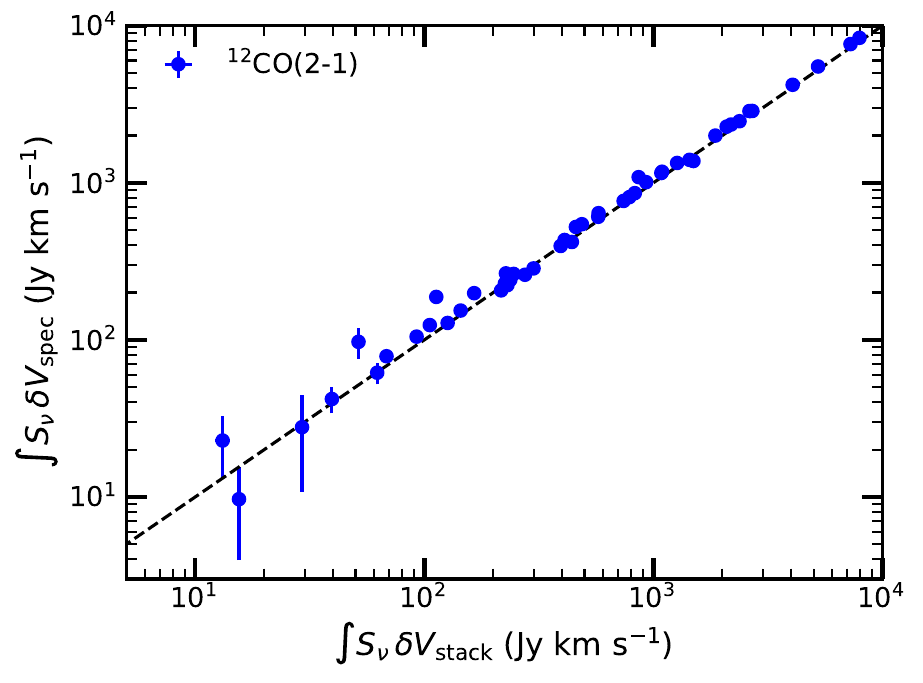}\\
	\caption{\co(2-1) integrated line intensities estimated from our \stack\ analysis are shown on the x-axis, and compared with those measured from the total integrated galaxy spectra on the y-axis. A one-to-one correlation is indicated with a black dashed line, as a guide to the eye. Errors in the stacked fluxes are included, but are generally smaller than the plotting symbol. The \co(2-1) fluxes from \stack\ compare well with those estimated in \protect  \cite{Brown2021}. }
	\label{fig:fluxcomp}
\end{figure}

As a cross-check, Figure \ref{fig:fluxcomp} shows the \co(2-1) flux retrieved from our stacking analysis, versus that measured by \cite{Brown2021} by standard summation of the observed \co(2-1) line profile. The integrated intensities closely follow the one-to-one black dashed line, suggesting our analysis is successful in retrieving all the flux. A few individual galaxies scatter low (i.e. having lower integrated intensities in the stacks than those derived from pure integration. These tend to be those systems that (as discussed in \citealt{Brown2021}) have larger negative sidelobes still present in the ACA data, and thus the true integrated intensity uncertainties are likely underestimated (both in \citealt{Brown2021} and here).

\subsection{\co/\tco\ global line ratios}

As discussed above, if one makes the common assumption that \tco\ in our galaxies is optically thin, then the \co/\tco\ ratio contains information both about the optical depth of the \co\ emitting gas ($\tau_{12}$) and abundance variations as:

\begin{equation}
\frac{\mathrm{^{12}CO}}{\mathrm{^{13}CO}} \propto \frac{[\mathrm{^{12}C/^{13}C}]}{\tau_{12}}.
\label{optdepth_eqn}
\end{equation}
 
In principle the observed line ratio also could depend on the temperature of the gas, especially as here we are not observing the ground state. However, the excitation temperatures of the 2--1 transition of \co\ and \tco\ are very similar (both $\approx10$\,K), and thus  temperature dependencies are unlikely to play a major role.

When integrating over entire galaxies, the \co/\tco\ ratio found in typical field spirals is $\approx10$ \citep[e.g.][]{2019ApJ...880..127J,2023ApJS..268....3C}, with values up to $\gtsimeq$20 found in some more extreme starburst systems \citep[see e.g.][]{2014MNRAS.445.2378D,2019ApJ...879...17B}. With an assumed $[\mathrm{^{12}C/^{13}C}]$ of 70 \citep[e.g.][]{1993A&A...274..730H}, this implies that the \co\ emitting gas is quite optically thick in these systems ($\tau_{12}=$3.5 -- 7).

 The  \co/\tco\ values we find here for our cluster spiral galaxies (listed in Table \ref{Table2}) are more extreme, ranging from 10 -- 43 (see Section \ref{gassfrden}), with some lower limits being even higher (the most extreme being $\gtsimeq62$ in one object). These extreme ratios are more similar to those found in luminous infrared galaxies and are unlikely to result from errors in the stacking (as the \co\ data used to stack is of high quality, and the isotopologue emission is expected to share its kinematics). This suggests we may already be seeing signs of the cluster environment affecting conditions within the molecular gas. In the following sections we explore how the  \co/\tco\ ratio varies with galaxy properties to attempt to explain what is driving these differences. These galaxy properties are taken directly from the other VERTICO papers \citep{Brown2021,2022ApJ...933...10Z, 2022arXiv221116521J, 2022ApJ...940..176V,Watts2023}.
We find no clear evidence of  \co/\tco\ variation with stellar mass, the cluster-centric distance, galaxy inclination, or with the radial extent of the gas \citep[as measured in][]{Brown2021}. 

\subsubsection{Gas and star formation surface density correlations}
\label{gassfrden}
\begin{figure}
	\includegraphics[width=0.45\textwidth,trim=0cm 0cm 0cm 0cm,clip]{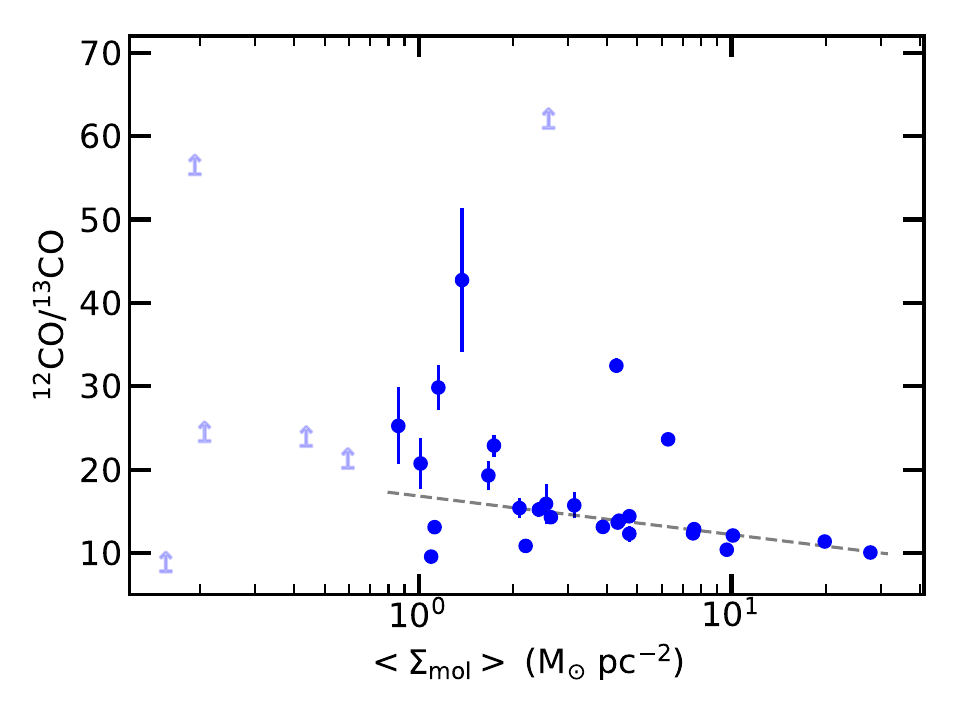}
	\caption{The global \co(2-1)/\tco(2-1) ratio for our VERTICO sources, plotted as a function of the mean molecular gas surface density in each galaxy. Galaxies with detections in both lines are shown as blue points with associated error bars, and blue arrows denote lower limits (i.e. where the   \co(2-1)/\tco(2-1) line ratio was not constrained at $5\sigma$). Uncertainties in the mean gas surface density measurements are plotted, but are too small to be seen (and are dwarfed by the systematic $X_{\rm CO}$ uncertainty, which is not plotted). As a guide to the eye we show with a dashed line the best-fitting linear correlation for our detected points.	 A significant (Spearmans rank correlation coefficient $\rho=-0.56$, $p$-value of 0.002) negative trend is present, with higher surface density systems showing lower \co(2-1)/\tco(2-1) ratios.}
	\label{fig:siggas_c12c13}
\end{figure}

In Figure \ref{fig:siggas_c12c13} we show the \co/\tco\ ratio, plotted as a function of the mean molecular gas surface density within the molecular gas disc of each object. This is measured from the VERTICO product database, calculating the arithmetic mean of the molecular gas surface density of all detected pixels.  Galaxies with high average column densities of molecular material are found to have lower \co/\tco\ ratios, and vice versa. In almost all cases, the galaxies that are not detected in \tco\ are those with extremely low average \co\ column densities. 
This correlation is significant- with a Spearmans rank correlation coefficient (for the detected galaxies) of $\rho=-0.56$, and a $p$-value of 0.002. As a guide to the eye we show with a dashed line the best-fitting linear correlation for our detected points, fitted using the \textsc{LTSFIT} package of \cite{2013MNRAS.432.1709C}.  This best-fitting line has the equation

\begin{equation}
\frac{\rm ^{12}CO}{\rm ^{13}CO}= (-4.6\pm0.5)\log_{10}\left(\frac{<\Sigma_{\rm mol}>}{\mathrm{M}_{\odot}\,\mathrm{pc}^{-2}}\right) + (16.8\pm0.5).
\end{equation}

\begin{figure}
	\includegraphics[width=0.45\textwidth,trim=0cm 0cm 0cm 0cm,clip]{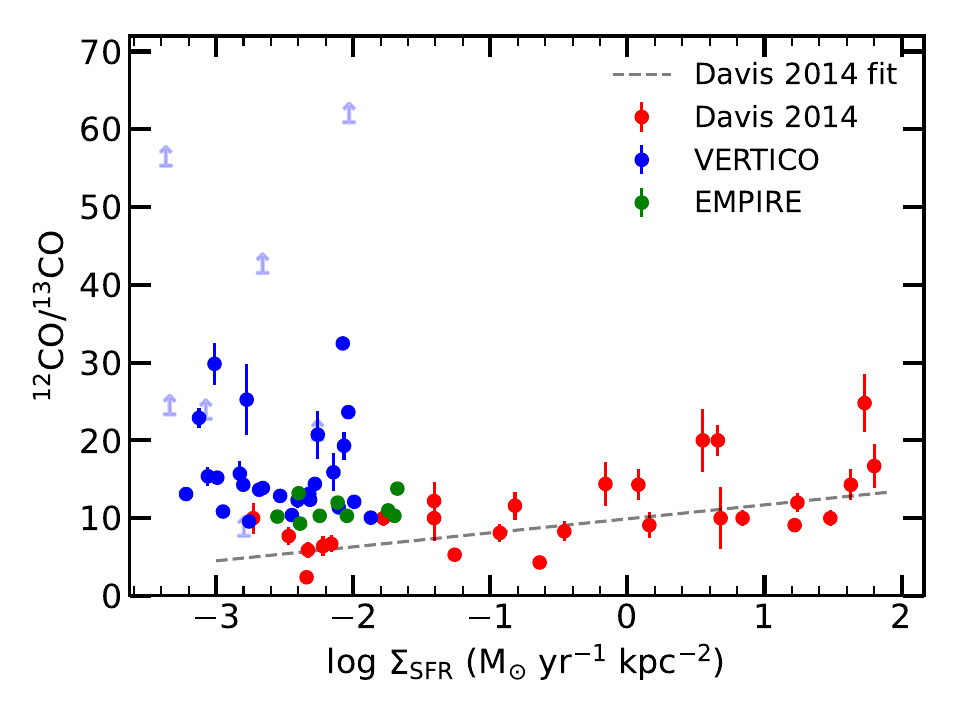}
	\caption{The global \co(2-1)/\tco(2-1) ratio, plotted as a function of the mean star-formation rate surface density. The VERTICO points are shown as dark blue circles, with lower limits indicated with arrows. Green and red points shown compiled literature data from EMPIRE \protect \citep{2019ApJ...880..127J} and \protect \cite{2014MNRAS.445.2378D}, respectively. The best-fit for the  \protect \cite{2014MNRAS.445.2378D} sample is shown as a grey dashed line. The VERTICO galaxies have higher global \co(2-1)/\tco(2-1) ratios than both the EMPIRE and \protect \cite{2014MNRAS.445.2378D} spiral (e.g. log star-formation rate surface density $<-1.5$) galaxies.  }
	\label{fig:SigmaSFR_c12c13}
\end{figure}

Given that a correlation is present between gas surface density and isotopologue line ratios, it is instructive to consider if a similar relationship exists with star formation rate surface density. This parameter is calculated in an analogous way to the mean surface density, calculating the arithmetic mean of the star-formation rate surface density for all detected pixels.  Figure \ref{fig:SigmaSFR_c12c13} shows the VERTICO data in dark blue, along with data from EMPIRE (green points), and the data compiled by \cite{2014MNRAS.445.2378D} in red. Unlike the \cite{2014MNRAS.445.2378D} data (where a positive trend is seen between the star formation rate surface density and  \co/\tco; seen also at \co(1-0) with more galaxies in \citealt{2023ApJS..268....3C}),  {the VERTICO galaxies show no significant correlation with star formation rate surface density (Spearmans rank correlation coefficient $\rho=-0.11$, $p$-value of 0.6)}. The VERTICO objects are clearly less star forming than the starbursts included in the \cite{2014MNRAS.445.2378D} sample, but given their \co/\tco\ line ratios are significantly higher than the spiral galaxies from that work (and than the spirals in e.g. EMPIRE, which are reasonably well mass matched to those in our sample) some other mechanism is clearly also at play. This is discussed further in Section \ref{discuss}.

\subsection{Correlations with morphology}

In this paper we have focused our study on the population of spiral galaxies in the Virgo cluster. In order to determine if the cluster environment changes the conditions within the gas it is crucial that we also compare to field samples. In addition, a large fraction of galaxies in the cluster are early-type galaxies (ETGs). These systems have likely been in the cluster longer, and have thus been even more strongly affected by environment. We thus include comparisons to these systems here. We note that the comparison samples available are small, and thus we cannot formally match the properties of the VERTICO galaxies to the late-type systems in the field. This caveat is discussed further in Section \ref{sec:etg_vs_ltg}.

\begin{figure}
	\includegraphics[width=0.45\textwidth,trim=0cm 0cm 0cm 0cm,clip]{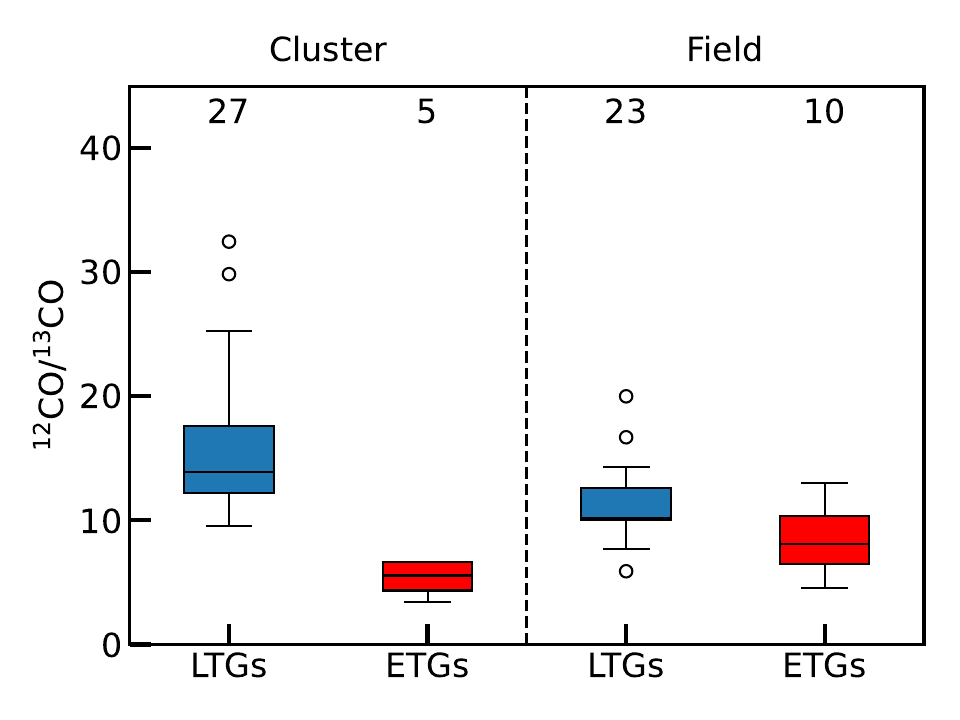}\\
	\caption{Box and whisker plots showing the distribution of \co/\tco\ line ratios in the Virgo cluster and field, for both Late-type galaxies (LTGs, e.g. spirals)  and early-type galaxies (ETGs). Coloured boxes (blue LTGs, red for ETGs) show the interquartile range, with a horizontal line at the median line ratio. Outlier galaxies that lie beyond the outer whiskers are shown as open circles.  The bars show the range of the data, with outliers shown as open circles. The underlying data is drawn from this work, EMPIRE and the \atlas\ survey \protect \citep{2015MNRAS.450.3874A}. LTGs have higher line ratios in the cluster than the field, while the trend is reversed for ETGs.}
	\label{fig:env_wetgs}
\end{figure}

In Figure \ref{fig:env_wetgs} we show the \co/\tco\ line ratios of Virgo cluster spirals (from this work), and compare these with field spirals from EMPIRE, and field and Virgo cluster ETGs from the \atlas\ survey \citep[][as introduced in Section \ref{intro}]{2019ApJ...880..127J,2015MNRAS.450.3874A}. The VERTICO systems cover a similar range of stellar mass  to the ETGs (9.5$<$log(M$_{*}$/M$_{\odot}$)$<$11.5), but are not directly matched in this parameter. 

The VERTICO cluster spiral galaxies have higher average \co/\tco\ line ratios than those found in field systems (although their distributions overlap). 
While the number of ETGs observed in the cluster is small, it seems that ETGs in the field have higher \co/\tco\ line ratios than those in the cluster, reversing the trend found in spirals. 
These differences between the median \co/\tco\ in each galaxy environment/type group are statistically significant (see Table \ref{ad_table} for $k$-sample Anderson-Darling test results), despite the low number of systems observed. The cause of this effect is discussed further in Section \ref{sec:etg_vs_ltg}.

\begin{figure}
	\includegraphics[width=0.48\textwidth,trim=0cm 0cm 0cm 0cm,clip]{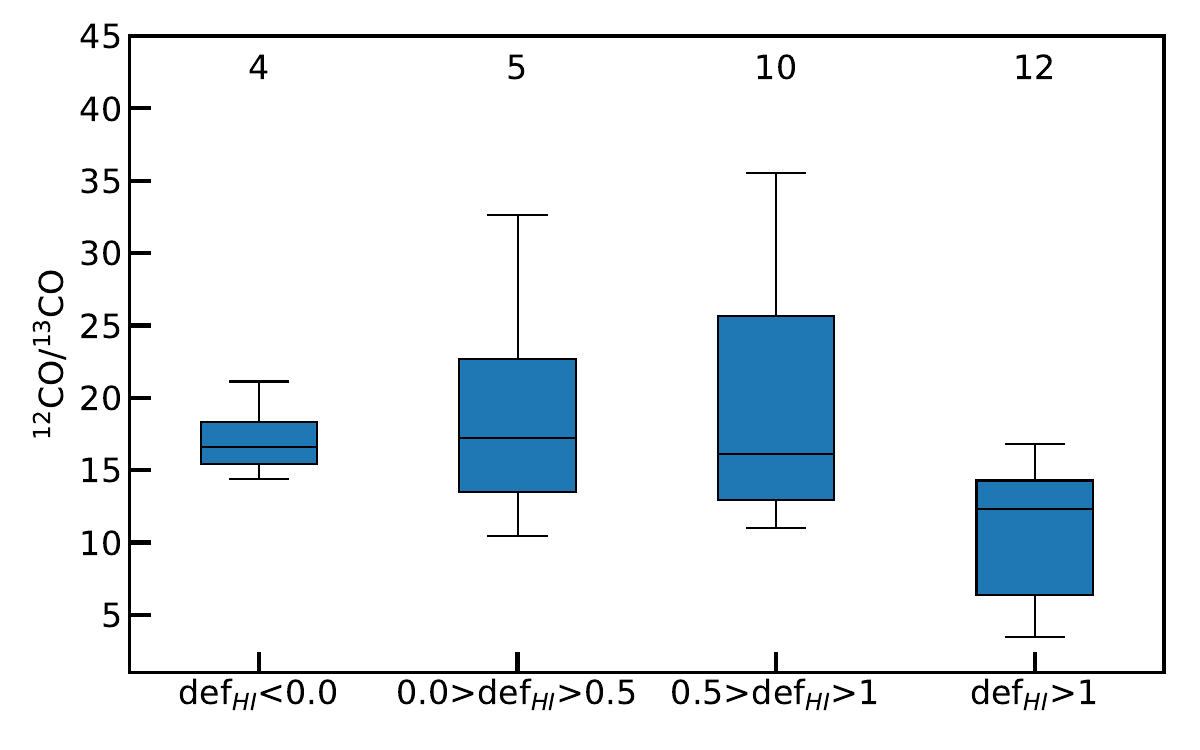}
	\caption{{Box and whisker plots showing the \co/\tco\ line ratios of early- and late-type cluster galaxies grouped as a function \hi-deficiency. The number of objects included in each box is indicated along the top of each plot. Blue boxes show the interquartile range, with a horizontal line at the median line ratio. Outlier galaxies that lie beyond the outer whiskers are shown as open circles. Galaxies with the highest \hi-deficiency (def$_{\rm HI}>1$) have the lowest \co/\tco\ line ratios, and galaxies with intermediate deficiencies show the largest scatter.}}
	\label{fig:hi_def}
\end{figure}

\subsection{Correlations with \hi\ deficiency}
{Given the above result, it is instructive to compare  how the environmental mechanisms identified in the more extended \hi\ gas affect the isotopologue line ratios in cluster galaxies of all morphologies. Various works in the VERTICO series have used the \hi\ deficiency \citep{Yoon2017}  to quantify the amount of \hi\ each system has with respect to field environments, and thus the strength/duration of environmental effects each galaxy has been subjected to. }

 {Figure \ref{fig:hi_def} shows box-and-whisker plots of \hi\ deficiency for all cluster galaxies (both early-and late-type), split by \hi-deficiency (where the deficiencies are taken from \citealt{Yoon2017} and \citealt{2012MNRAS.422.1835S}). The boxes show the median and interquartile ranges, and the bars the range of the data (with open circles showing outliers). The number of systems within each bin is shown at the top of the plot.
 The most \hi-deficient galaxies have the lower \co/\tco\ line ratios. While this bin contains the majority of early-type systems, the same trend appears to hold in just the late-type galaxies, albeit with lower number statistics.  
These differences are statistically significant (see Table \ref{ad_table} for $k$-sample Anderson-Darling test results).}  The scatter appears to increase for galaxies at intermediate deficiencies, although again our statistics are limited.
 {This suggest an environmentally driven trend, with galaxies of all morphological types that have higher \hi-deficiencies (i.e. those that are likely to have been in the cluster longer) having lower line ratios.} This will be discussed further in Section \ref{sec:etg_vs_ltg}.

\begin{table}
\caption{$k$-sample Anderson-Darling test results for variation of line ratios with galaxy type, environment and \hi-deficiency.}
\centering
\begin{tabular}{lrr}
\hline
Test Performed & Statistic & $p$-value \\
(1) & (2) & (3)\\
\hline
Cluster Spirals vs Cluster ETGs & 10.769 & 0.001\\
Field Spirals vs Field ETGs & 2.615 & 0.027\\
Field Spirals vs Cluster Spirals & 8.416 & 0.001\\
Field ETGs vs Cluster ETGs & 2.673 & 0.026\\
\hline
{def$_{\rm HI}<0$ vs def$_{\rm HI}>1$} & 3.764 & 0.012\\
{$0<$def$_{\rm HI}<1$ vs def$_{\rm HI}>1$} & 4.198 & 0.007\\
\hline
\end{tabular}
\\ \textit{Notes:} Column 1 describes the test performed. Column 2 contains the normalized k-sample Anderson-Darling test statistic, and Column 3 the associated $p$-value. \label{ad_table}
\end{table}

\begin{figure*}
	\includegraphics[width=0.8\textwidth,trim=0cm 0cm 0cm 0cm,clip]{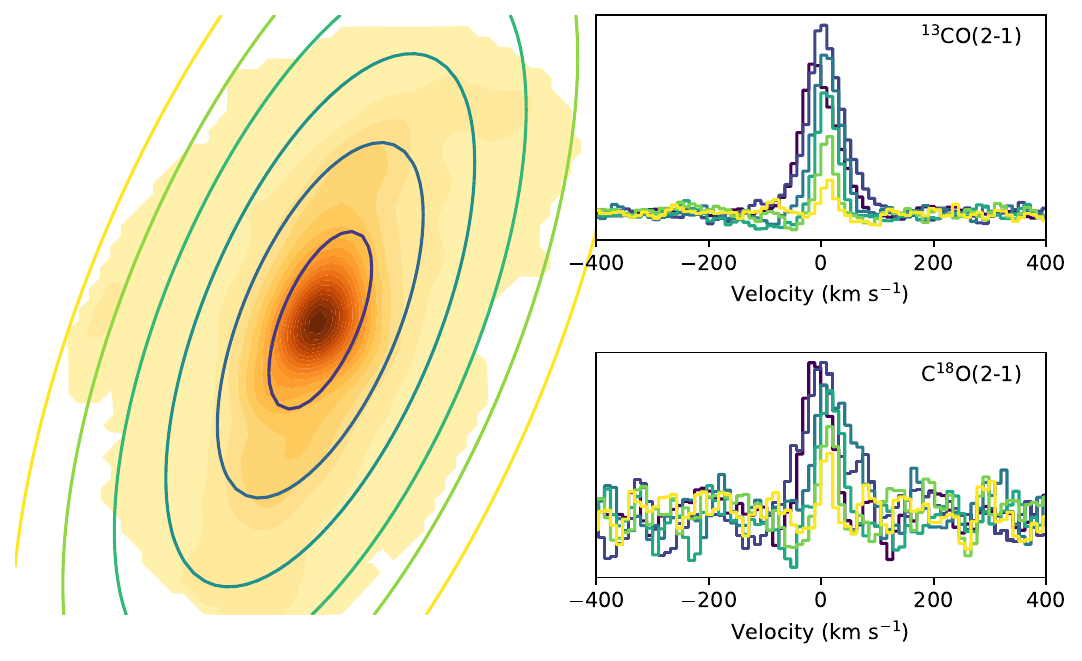}\\
	\caption{\tco(2-1) and \ceo(2-1) stacked spectra extracted from beam-width radial bins in VERTICO galaxy NGC4568. The VERTICO integrated intensity map of this galaxy is shown on the left of the figure, with coloured ellipses indicating the stacking apertures. The \tco(2-1) and \ceo(2-1) stacked spectra in the right hand plots (presented with an arbitrary y-scale for visibility) are plotted using the colour corresponding to the aperture from which they were extracted. In this galaxy each line was well detected in each radial aperture. We note that this object has a close companion (NGC4567), which has been removed by masking the portion of the cube containing its emission. }
	\label{fig:advert_radial_stack}
\end{figure*}

\begin{figure*}
	\includegraphics[width=1\textwidth,trim=0cm 0cm 0cm 0cm,clip]{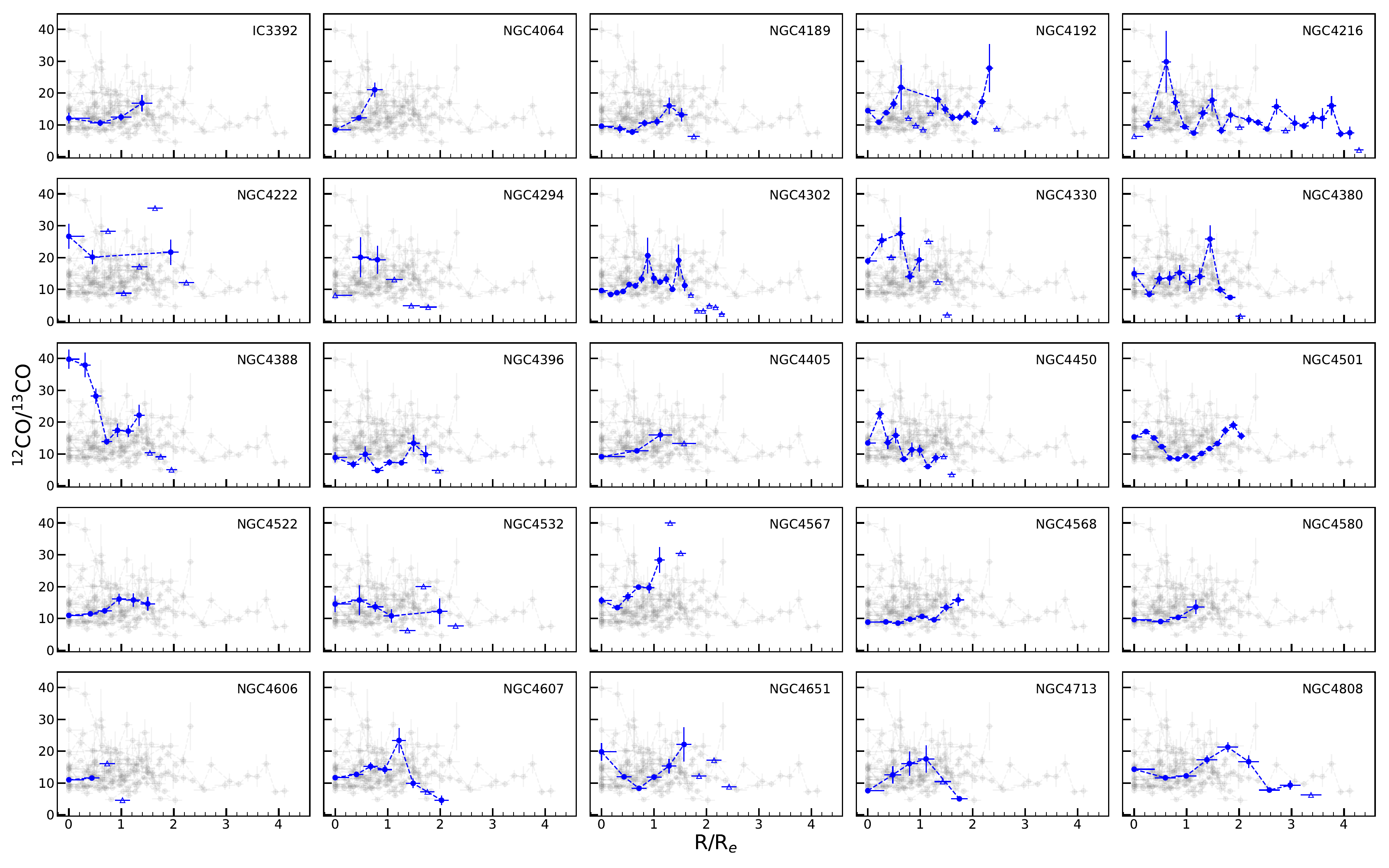}\\
	\caption{ \co/\tco\ line ratio radial profiles, in the 25 galaxies where they could be constrained in at least two radial bins. The solid blue points in each panel, connected by dashed lines, show the line ratios derived in each detected radial bin. Open triangles denote 3$\sigma$ upper limits. The ensemble of all radial points from all galaxies is shown in light grey in each panel as a guide to the eye.}
	\label{fig:radial_stack_r21}
\end{figure*}

\subsection{Radial profiles}
\label{sec:radprof}

Having studied in detail the global line ratio correlations, we now proceed to stack the data within beam-width sized radial apertures in each system, allowing us to study radial gradients in the isotopologues. Figure \ref{fig:advert_radial_stack} shows an example of this radial stacking for NGC4568. 

Figure \ref{fig:radial_stack_r21} shows the \co/\tco\ line ratios as a function of radius (normalised by the $r$-band effective radius of each system) in the 25 galaxies where they could be well constrained. The blue points in each panel, connected by dashed lines, show the line ratios derived in each radial bin for each system. The ensemble of all radial points from all galaxies is shown in light grey in each panel as a guide to the eye. A similar plot for the galaxies with constrained \co/\ceo\ profiles is shown in the Appendix (Figure \ref{fig:radial_stack_rco_ceo}). 
The majority of systems have very flat line-ratio gradients, with small systematic deviations present in certain systems (e.g. the centre of NGC4388, and an area around $\approx$1.8\,R$_{\rm e}$ in NGC4808). This will be discussed further in Section \ref{sec:radial_discuss}.

\subsection{Tailed galaxies}

Some of the galaxies in our sample have \hi\ tails that suggest they are being actively environmentally processed. Such systems may have different ISM conditions in the tail (where the gas would be expected to be more diffuse) versus the leading edge of the system (where e.g. tidal and ram-pressure stripping would be expected to compress the gas). 

To search for such a signature we stack quadrants of each galaxy towards and away from the direction of the \hi\ tail, using the tail angles derived in \cite{2022MNRAS.517.2912L}. We were able to complete this analysis for five galaxies in our sample. As tabulated in Table \ref{tail_table}, three (NGC4299, NGC4396 and NGC4501) show clear enhancement in the \co/\tco\ line ratios in quadrant of the galaxy in the direction of their tails. One galaxy, NGC4302 shows no change in the line ratio towards or away from the nucleus, and one system (NGC4351) shows a lower  \co/\tco\ ratio in the direction of its tail. These results will be discussed further in Section \ref{sec:etg_vs_ltg}.

\begin{table}
\caption{\co/\tco\ values towards and away from \hi\ tails.}
\centering
\begin{tabular}{lrrr}
\hline
 Galaxy & Tail direction & Away from tail & Enhancement  \\
 &  \co/\tco$_{\rm tail}$ &  \co/\tco$_{\rm lead}$ & Tail/ Lead \\

(1) & (2) & (3)\\
\hline
NGC4299 & $>$32.8 & 4.2 $\pm$ 0.9 & $>$7.9 \\
NGC4302 & 11.9 $\pm$ 0.6 & 11.9 $\pm$ 0.4 & 1.0$\pm$ 0.06 \\
NGC4351 & 14.4 $\pm$ 2.3 & $>$52.3 & $<$0.3 \\
NGC4396 & $>$14.6 & 9.8 $\pm$ 1.3 & $>$1.4\\
NGC4501 & 14.1 $\pm$ 0.4 & 11.8 $\pm$ 0.2 & 1.2 $\pm$ 0.04 \\
\hline
\end{tabular}
\\ \textit{Notes:} Column 1 lists the galaxies. Column 2 and 3 list the \co/\tco\ ratios in a specific direction (towards or away from the \hi\ tail of the system). Where \tco\ is not detected these values are calculated using 3$\sigma$ upper limits, making the resulting ratio a lower limit. Column 4 contains the fractional enhancement of \co/\tco\ in the tail direction e.g.  (\co/\tco$_{\rm tail}$)/(\co/\tco$_{\rm lead}$). Three systems have clear enhancements in the tail direction, while one shows no change, and one shows a decreased line ratio in the tail direction. \label{tail_table}
\end{table}

\subsection{Comparison of \tco\ and \ceo}
\label{tcoandceo}
\begin{figure}
	\includegraphics[width=0.48\textwidth,trim=0cm 0cm 0cm 0cm,clip]{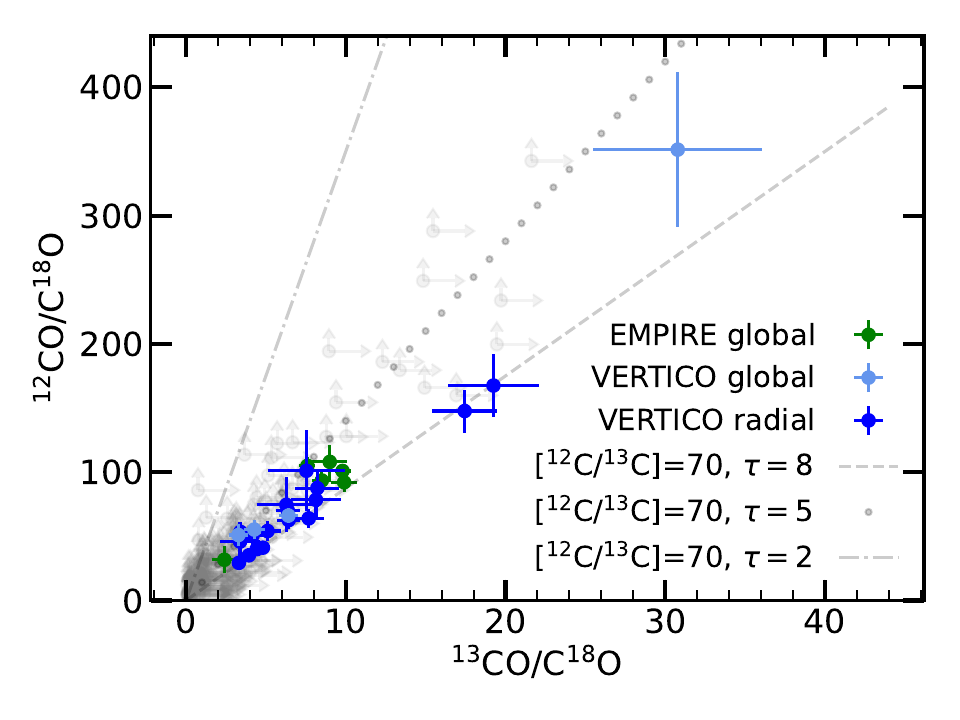}
	\caption{The  \co/\ceo\ line ratio plotted as a function of the \tco/\ceo\ line ratio. Radial bins where both lines were observed and detected are shown as dark blue circles with error bars, while global measurements for detected galaxies are shown in lighter blue. Lower limits are shown as grey points with limit arrows. Global measurements of galaxies from the EMPIRE \protect \citep{2019ApJ...880..127J} survey are shown in green. As expected, a positive trend is seen in the detected systems, whereby galaxies with lower \co/\ceo\ line ratios also have lower \tco/\ceo\ line ratios. The trends expected if \co\ has an optical depth $\tau$ of 2, 5 and 8 and [$^{12}$C/$^{13}$C]=70 are shown as grey dot-dashed, dotted and dashed lines, respectively. If \tco\ and \ceo\ are both optically thin then [$^{18}$O] abundance variations would move points along these lines.}
	\label{fig:c18o_c13o}
\end{figure}

We detect the weaker \ceo\ line through global stacks in eight of our galaxies (although due to the use of archival data only four of these systems were also observed in \tco). Radial stacking allows us to detect \ceo\ in individual regions in twelve galaxies. Comparison of the bins with detected \ceo\ in galaxies where the line was not detected in the global stack show that the total fluxes of the lines are consistent with the derived global upper limits in Table \ref{Table1}. This suggests local \ceo\ enhancements are present which can be missed in global stacks. Here we compare  \tco\ and \ceo\ line ratios (listed in Table \ref{Table2}) both globally and in radial bins.  
Figure \ref{fig:c18o_c13o} shows the  \co/\ceo\ line ratio plotted as a function of the \tco/\ceo\ line ratio. Radial bins where both lines were observed and detected are shown as dark blue circles with error bars, while global measurements for these same galaxies are shown in lighter blue. Lower limits are shown as grey points with limit arrows. Global measurements of galaxies from the EMPIRE survey are shown in green. Theoretical lines of constant \co\ optical depth from Equation \ref{optdepth_eqn} (assuming a \tco-to-\co\ abundance of 70; e.g. \citealt{1993A&A...274..730H}) are shown in grey, to guide the eye.

A strong positive trend is observed in the detected points, whereby galaxies with larger \co/\ceo\ line ratios also have larger \tco/\ceo\ line ratios. While these axes are correlated, and thus uncertainties in the weaker \ceo\ line will induce a correlation, the spread of points is larger than can be explained through this alone. This suggests real [$^{13}$C/$^{18}$O] abundance changes (or some type of isotope dependent fractionation) may be present. This will be discussed further in Section \ref{discuss}.

\section{Discussion}
\label{discuss}
In this work we have attempted to detect the CO isotopologues \tco\ and \ceo\ both globally, and across the discs of a sample of 48 spiral galaxies in the Virgo cluster.  
We mapped at least one isotopologue in seventeen sources, and detected \tco\ globally via stacking in 30/48. We have shown above that the isotopologue line ratios correlate with galaxy properties in a variety of ways. We discuss below the potential physical drivers of the observed correlations.

\subsection{The spatial distribution of isotopologues}
\label{discuss_directdetect}
\label{sec:radial_discuss}
\subsubsection{\tco}
In some galaxies where the gas is particularly bright we detect \tco\ emission directly in the image plane. As shown in Figures \ref{fig:directdetect} and \ref{fig:directdetect_appendix}, where \tco\ is detected directly it tends to be found in the regions which also have bright \co\ -  the centres of the galaxies, and in the densest knots within the spiral arms. 
Through stacking we are able to detect emission in almost all galaxies with $<\Sigma_{\rm gas}>\gtsimeq 1\,$\msun\,pc$^{-2}$. The only exception to this is NGC4383, a galaxy that is very \hi-rich and quite dynamically disturbed, and hosts a large ionised gas outflow \citep{2024MNRAS.530.1968W}. We are able to set strong upper-limits on the \tco\ emission from this galaxy (\co/\tco\ $\gtsimeq$63). Taken at face value this would suggest that the \co\ emission in this system is nearly optically thin, or that [$^{12}$C/$^{13}$C] is very different from the value typically assumed ($\approx$70).

In the majority of our galaxies the abundance and optical depth of \co\ and \tco\ seems to vary little throughout the disc (at least on $\approx$600\,pc scales). This is supported by the resolved radial profiles of our systems in Figure \ref{fig:radial_stack_r21}. The \co/\tco\ line ratios typically vary little with radius (less than a factor of two in almost all cases), suggesting \tco\ is well distributed without significant variation in abundance or gas conditions within our galaxies.  A few systems have enhancements in their \co/\tco\ ratio at certain radii, which seem to correspond to areas where significant gas flows/disturbances are expected (e.g. the nuclear disc within the bar of NGC4388, and the lopsided component in NGC4808).

The surveys of \cite{2018MNRAS.475.3909C,2022A&A...662A..89D} and \cite{2023ApJS..268....3C} show that nearby field spiral galaxies have similar \co/\tco\ line ratios to those we find here, albeit they sometimes show (very) mild positive radial gradients in this ratio which we do reproduce. Cluster galaxies are typically more metal rich (and these metals are better mixed) due to the time they have spent in the cluster \citep[e.g.][]{2009MNRAS.396.1257E,2021ApJ...923...28F,2022A&A...660A.105L}, or because the metal poor gas mostly located in the outer disc is preferentially removed by environmental effects \citep{2013A&A...550A.115H}. Thus if the \co/\tco\ gradients in field spirals are partially due to abundance variations it is expected that they would be weaker in cluster objects. Alternatively, the high ram- and hydrostatic mid-plane pressure experienced by the molecular gas in cluster galaxies at some stages of their lives 
may compress the molecular clouds, and thus increase the optical depth of the \co\ in the outer discs of our systems, again flattening the \co/\tco\ gradients. 

\subsubsection{\ceo}

A somewhat different picture is present in \ceo. In the two of the three galaxies where we detect \ceo\ directly it avoids the brightest \co\ peak at the galaxy centre, and instead is more common further out in the dense knots within the galaxy disc. The radial profiles seen in Figure \ref{fig:radial_stack_rco_ceo} show a range of behaviours. While we only have a small sample, and low S/N, a similar effect is seen in several of the galaxies studied by \cite{2017ApJ...836L..29J}, suggesting this could be a physical effect, perhaps be due to abundance variations, or changing conditions within our galaxies. Given the lack of radial gradients in \co/\tco\ we consider abundance variations more likely, and address the possibility in more depth below. 

\subsection{Isotopologue abundances}

Figure \ref{fig:c18o_c13o} compares the \co/\ceo\ and \tco/\ceo\ line ratios, and suggests that real abundance variations are likely present in our cluster galaxy population. Both \tco\ and \ceo\ are likely to be optically thin (with \tco\ typically only becoming optically thick in the densest star forming regions, which do not dominate the emission we detect on global scales). Given this, the variations we observe (from 2 \ltsimeq \tco/\ceo\ \ltsimeq 30) are unlikely to derive from optical depth differences, and very likely reflect real changes in abundances.  

In Figure \ref{fig:c18o_c13o} both the spatially resolved and integrated data points from VERTICO and EMPIRE fall along a line of approximately equal mean [$^{12}$C/$^{13}$C]/$\tau$. 
This suggests that the $^{18}$O/$^{16}$O abundance ratio may vary more than the $^{13}$C/$^{12}$C abundance, although we would require additional isotopologue observations (e.g.  $^{13}$C$^{18}$O) to confirm this directly.

Such changes can arise because of several physical effects (that may, or may not relate directly to environmental quenching of these systems). 
Firstly, it is possible that the underlying abundances of these isotopes are the same in our sample galaxies, but the variations we observe are caused by selective photodissociation or isotope-dependent fractionation of the molecules. Both \tco\ and \ceo\ are preferentially dissociated by UV photons in the outer layers of molecular clouds, and due to their lower abundance are less protected than \co\ \citep[e.g.][]{1988ApJ...334..771V}. However these spiral galaxies are not highly star forming compared with e.g. the Milky Way, and span a relatively small range of star formation rate surface density. Thus it seems unlikely that such large changes in abundance can be caused by dissociation. 
Similarly, isotope-dependent fractionation can cause the heavier $^{13}$C and $^{18}$O isotopes to preferentially form molecules (and deplete onto dust grain icey mantles) at the lowest temperatures \citep[e.g.][]{1976ApJ...205L.165W}. However, these temperatures are only reached deep within the star forming cores of molecular clouds. Here we study the bulk molecular medium on kiloparsec scales where the diffuse medium likely dominates. 
Thus it seems unlikely that isotope-dependent effects are dominating what we see.

The second possibility is that the line ratio differences we see could arise due to real variations in the abundances of the isotopes (i.e. selective nucleosynthesis). Recent work has highlighted that the $^{12}$C, $^{13}$C and $^{18}$O isotopes are produced by different mass stars, and thus the amount of each available to be locked up in molecules depends on both time, and the input stellar population \citep[e.g.][]{2017MNRAS.470..401R,2017ApJ...840L..11S,2018Natur.558..260Z,2019ApJ...879...17B}.  Interpreted naively, the work of e.g.  \cite{2024arXiv240505317G} would suggest a bottom heavy stellar initial mass function would be required to explain the high \tco/\ceo\ ratios we detect. However, we caution that our cluster spiral galaxies are very different environments than the high-redshift starburst galaxies studied with this technique to date, and have much more complex star formation histories. We further caution that $^{18}$O is  a challenging isotope to predict, which can be impacted by stellar physics including stellar rotation and the impact of novae \citep{2003MNRAS.342..185R,2019MNRAS.490.2838R}. While we hesitate to draw strong conclusions from this data, these cluster systems present an exciting new laboratory for exploring the IMF sensitivity of molecular tracers further.

\subsection{Environmental influences and gas processing}
\label{sec:etg_vs_ltg}

The isotopologue line ratios provide us with a window into the conditions within the ISM of each galaxy. We can use this tool to study the impact of the cluster environment on the ISM. 
We have seen that the isotopologue line ratios are significantly different in cluster galaxies compared to field systems, and that this trend is different in early- and late-type galaxies. {Across galaxy types within the cluster the line ratio correlates with \hi-deficiency, a rough proxy for time spent within the cluster.}

\begin{figure}
	\includegraphics[width=0.48\textwidth,trim=0.25cm 0.25cm 0.2cm 0cm,clip]{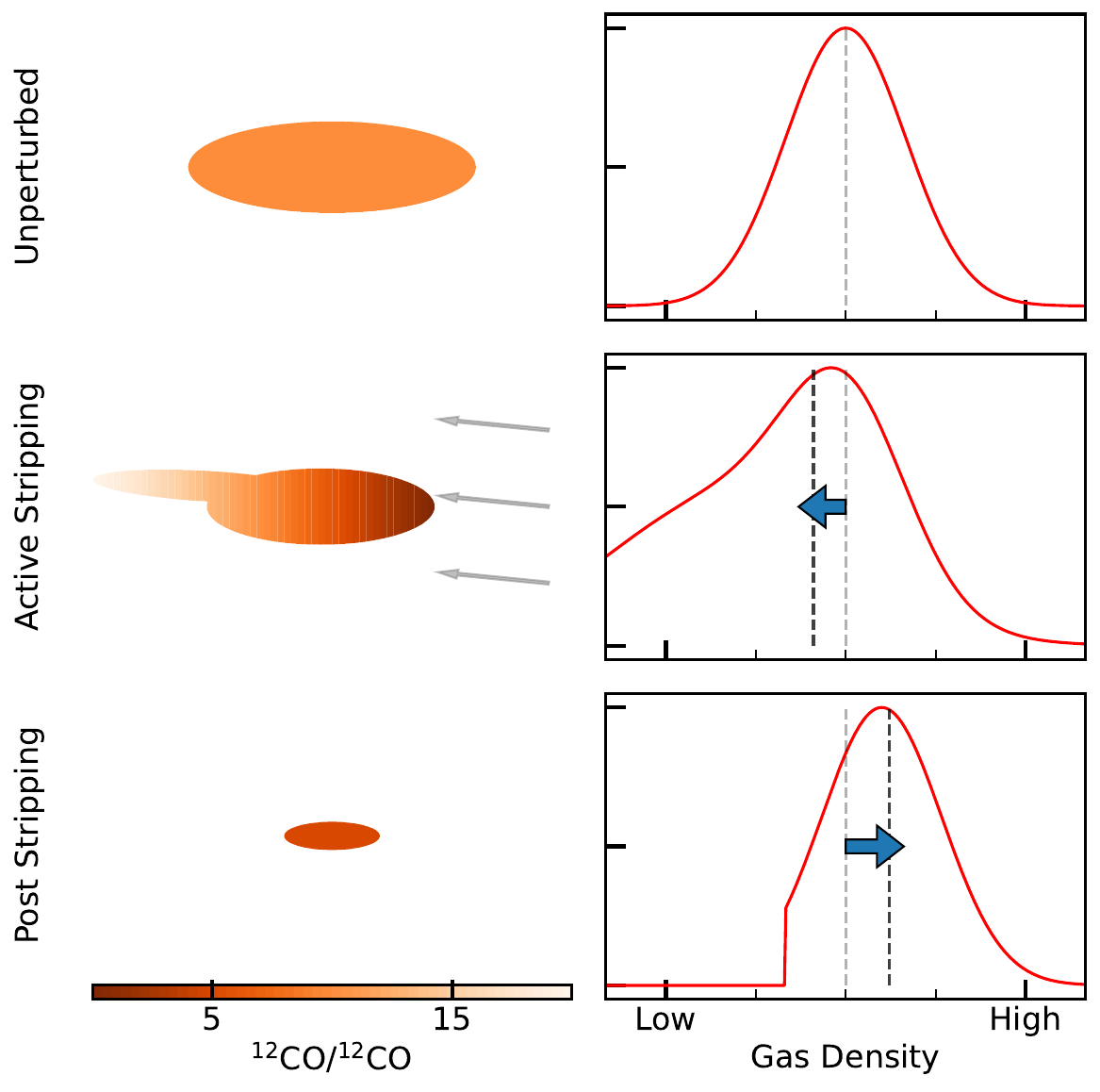}
	\caption{A cartoon illustrating the environmental processing scenario outlined in Section \ref{sec:etg_vs_ltg}. The first column shows the gas in the galaxy, coloured by its \co/\tco\ ratio. The second column shows the gas density probability distribution function (PDF) within this gas disc. The mean gas density of the unperturbed galaxy is shown as a grey dashed line on each PDF, with a darked dashed line showing the resultant mean density after stripping.  As a galaxy is environmentally processed the material along the leading edge is compressed, while the material in the process of being stripped is at lower densities, reducing the mean \co\ optical depth overall. Once all the low density material has been removed by environmental processes, the remaining material is more centrally concentrated and at higher pressure, increasing the mean density and \co\ optical depth. }
	\label{fig:gasproc_cartoon}
\end{figure}

Here we suggest that a single scenario could potentially explain these trends. As galaxies fall into the cluster their ISM is processed and disturbed. Once they have been in the cluster a long time, however, the remaining gas settles deep within the galaxy core where it can no longer be disturbed. If these processes impact the mean optical depth of the gas (as has been suggested in interacting galaxies \citealt{2023ApJS..268....3C}) then this could explain the observed trends (see also \citealt{2021PASA...38...35C}). In Figure \ref{fig:gasproc_cartoon} we show a cartoon of this scenario, highlighting how the gas density probability distribution function (PDF) would be expected to change during the process, along with the \co/\tco\ line ratio.

This suggestion is further backed up by our analysis of the \co/\tco\ line ratio as a function {of \hi-deficiency (Figure \ref{fig:hi_def}). Galaxies with normal \hi-content have not yet been substantially affected by the cluster, and have a small range of \co/\tco\ line ratios. The systems with intermediate \hi-deficiencies show more scatter in their \co/\tco\ line ratios, likely because some of these are currently undergoing strong environmental processing, and as their gas is expelled, the average density (and thus average optical depth) drops (c.f. Figure \ref{fig:siggas_c12c13}). 
Systems with large \hi-deficiences have lower \co/\tco\ line ratios than either of the other classes. These galaxies are likely to have already had much of their ISM stripped, and what remains is a denser (higher optical depth and thus lower \co/\tco) core of the material that was more resilient to stripping.} This suggests that where environmental influences on the \hi\ gas are strong this can also disturb the denser molecular gas (in agreement with other analyses, both in this series; e.g. \citealt{2022ApJ...933...10Z,2022ApJ...940..176V,Watts2023}, and in the literature, e.g. \citealt{2018ApJ...866L..10L}).
The presence of these environmental effects likely overwhelms other correlations explaining why, for instance the \co/\tco\ ratios do not correlate the expected way with the star formation rate surface density (e.g. Figure \ref{fig:SigmaSFR_c12c13}). 

This scenario outlined above also explains the differing trends of \co/\tco\ line ratio as a function of environment for early- and late-type galaxies seen in Figure \ref{fig:env_wetgs}. A fraction of spiral galaxies in the cluster are currently being disturbed by environmental mechanisms, leading them to have higher \co/\tco\ line ratios on average than field LTGs. Early-type galaxies in the field have compact gas reservoirs that are typically denser than those found in LTG discs \citep[e.g.][]{2014MNRAS.444.3427D}, and thus have higher optical depths (and lower \co/\tco\ line ratios). In the cluster ETGs have typically been present for a long time (and they have high \hi-deficiencies) hence the environmental effect is exacerbated. {The \hi, and lower density molecular material} has been swept away long ago, leaving only the most dense, high optical depth molecular gas (see e.g. \citealt{2009ApJ...694..789T}). {These dense cloud cores would have low \co/\tco\ line ratios, similar to (but more extreme than) those seen in the \hi-deficient spiral galaxies.}

A large uncertainty is that only a small number of comparison objects have suitable \co(2-1) and \tco(2-1) observations. While larger samples of  \co(1-0) and \tco(1-0) observations are available (e.g. \citealt{2023ApJS..268....3C}), it is unclear if these can be directly compared without introducing additional uncertainties. While the scenario described above cannot be proven with the existing data (due to the degeneracy with abundance variations), it does explain well the observed correlations of the \co/\tco\ line ratio with galaxy properties. Further testing this scenario should be the focus of future observational (and theoretical) work.

{Another example of environmental processing affects is that in some (but not all) cases the molecular gas reservoirs aligned towards and away from \hi\ stripped tails have significantly different isotopologue line ratios than the bodies of their galaxies.
Three of our systems show increased \co/\tco\ in the direction of their tails with respect to the leading edge of the system, suggesting the disturbances caused by stripping or tidal forces are reducing the optical depth of the gas in the tail and/or compressing the material at the leading edge, increasing its optical depth. Some cases show no significant isotopologue differences, or decreases in line ratios however, so clearly the strength/presence of this effect may depend on the exact orbit and processes involved.}

\section{Conclusions}
\label{conclude}

In this work we studied the resolved and integrated properties of CO isotoplogue emission in the VERTICO sample of 48 spiral galaxies in the Virgo cluster.  We have shown that \tco\ is easily detected across the sample, both in the image plane for 35\% of the sample, and via stacking for nearly every galaxy where it was observed. This is despite the integration times of the observations being set to allow mapping of \co. \ceo\ is more difficult to detect directly, but is retrieved via stacking in $\approx$20\% of galaxies.  

We have then used these data to study trends with global galaxy properties, and how these isotoplogues vary across the faces of our galaxies. We have shown that the \co/\tco\ line ratio varies little with radius in our systems, either due to the high pressure in the cluster environment or due to enhanced chemical mixing. It does, however, vary significantly with a variety of galaxy properties, {including mean gas surface density, \hi-deficiency} and galaxy morphology.  {Some, but not all} galaxies with \hi\ tails also show \co/\tco\ line ratio variations towards and away from these tails.  \tco/\ceo\ line ratios vary significantly, both radially and between galaxies, suggesting variations in abundances are also present. We tentatively suggest these differences are larger in [$^{18}$O/$^{16}$O] than [$^{13}$C/$^{12}$C], although this will need to be confirmed with additional isotoplogue data (e.g. observations of $^{13}$C$^{18}$O). Such abundance variations may point to star formation history differences, or speculatively even IMF variation.   

We are able to find evidence (based on the \co/\tco\ observations) that the cluster environment is affecting the physical conditions within the molecular gas reservoirs of our systems. We present a scenario where environmental processes first decrease the density of the medium as material is stripped during first infall.  Once galaxies have been in the cluster for a longer period of time the tail of diffuse material has been removed, leading to a denser medium with higher optical depth, especially as systems transition to early-type morphologies. While this scenario is not unique, it does well explain the observations presented here, and those from related works.

The next steps in this type of analysis would include dedicated observations detecting even weaker isotopologues, such as $^{13}$C$^{18}$O, which would allow us to break the degeneracy between abundance variations and optical depth changes in our systems. Dedicated simulations including isotopologue yields would also allow us to disentangle the origin of the abundance changes we find, and how these are linked to the (pre-)processing and cluster specific processes underway in each galaxy. While such observations and simulations are challenging, they are feasible with the current facilities (ALMA and large supercomputers) and should allow us to make significant progress on understanding the impact of environment on the ISM, and thus quenching of galaxies.

\section*{Acknowledgements}

TAD acknowledges support from the UK Science and Technology Facilities Council through grants ST/S00033X/1 and ST/W000830/1. The research of CDW is supported by grants from the Natural Sciences and Engineering Research Council of Canada and the Canada Research Chairs program. NZ is supported through the South African Research Chairs Initiative of the Department of Science and Technology and National Research Foundation. IDR acknowledges support from the Banting Fellowship Program. V. V. acknowledges support from the ALMA-ANID Postdoctoral Fellowship under the award ASTRO21-0062. AC acknowledges support by the National Research Foundation of Korea (NRF), grants No. 2022R1A2C100298213 and No. RS-2022-NR070872. KS acknowledges funding from the Natural Sciences and Engineering Research Council of Canada (NSERC). MJJD acknowledges support from the Spanish grant PID2022-138560NB-I00, funded by MCIN/AEI/10.13039/501100011033/FEDER, EU.

\noindent This paper makes use of the following ALMA
data:\\
ADS/JAO.ALMA\#2019.1.00763.L, ADS/JAO.ALMA\#2017.1.00886.L,
ADS/JAO.ALMA\#2016.1.00912.S, ADS/JAO.ALMA\#2015.1.00956.S. 

\noindent ALMA
is a partnership of ESO (representing its member states), NSF (USA) and NINS
(Japan), together with NRC (Canada), MOST and ASIAA (Taiwan), and KASI
(Republic of Korea), in cooperation with the Republic of Chile. The Joint ALMA
Observatory is operated by ESO, AUI/NRAO and NAOJ. The National Radio
Astronomy Observatory is a facility of the National Science Foundation operated under cooperative agreement by Associated Universities, Inc

\section*{Data availability}
The data underlying this article is available from the ALMA archive using the project codes listed above. Processed data is available at the VERTICO website \url{https://www.verticosurvey.com/}.

\bsp	
\bibliographystyle{mnras}
\bibliography{bibMASSIVE_smbh.bib}
\bibdata{bibMASSIVE_smbh.bib}
\bibstyle{mnras}

\label{lastpage}
\clearpage
\appendix

\section{\textsc{Stackarator code description}}
\label{stack_appendix}

\textsc{Stackarator} is a tool for stacking datacubes of extended sources (such as nearby galaxies) to extract weak emission lines. It is written in the \textsc{python} coding language, and is publicly available at \url{https://github.com/TimothyADavis/stackarator}. Full installation instructions, and a tutorial notebook are available at that URL. 

\subsection{\textsc{Stackarator} Inputs}

The code requires two basic inputs. Firstly, a FITS datacube on which the stacking will be performed, 
\begin{equation}
D=f(\mathrm{RA}, \mathrm{Dec}, \mathrm{Vel}) \equiv f(\mathrm{RA}, \mathrm{Dec},\nu), 
\end{equation}
which can either be supplied with a velocity axis already in place, or with a frequency ($\nu$) axis which will converted to velocity using the rest-frequency ($\nu_0$) of the line you are attempting to stack. This doppler shift conversion uses the radio convention by default. I.e.
\begin{equation}
\mathrm{Vel}=c\frac{\nu_0-\nu}{\nu_0},
\end{equation}
where $c$ is the speed of light, and the other symbols are as defined above. The doppler shift convention can be changed by the user if required. From this datacube \stack\ will estimate the RMS sensitivity using the inner quarter of the cube in line-free channels specified by the user (by default channels 2 to 5 in the input cube). If the user wishes they can also supply the primary beam map to allow the code to create a 2D map of the RMS noise variation across the field, which will be factored into the uncertainty calculations. 

The second required input is a velocity field from some tracer which will be used to define the velocity shift when stacking each spaxel
 \begin{equation}
V=f(\mathrm{RA}, \mathrm{Dec}).
\label{velfield}
\end{equation}
This velocity field can be supplied either as a FITS file, or as a 2D array already in memory. This velocity field should have been produced using the same doppler shift conversion as specified when reading in the datacube (the radio convention by default).
 
The input velocity field is used to create an interpolation function, which uses linear spline interpolation to find the expected velocity of arbitrary new points, based on their RA and Dec values. This method has the advantage that the input velocity field doesn't need to be evaluated on the same grid as the input datacube. It could even be made from data at very different spatial/spectral resolutions (e.g. stacking CO based on \hi\ velocities), although this may impact the ability of the stack to detect the weakest lines if smaller-scale variations in velocity are present. If the input velocity field has had the galaxies' systemic velocity subtracted then \stack\ will take this into account, either using the \textsc{SYSVEL} keyword in the header, or a systemic velocity supplied by the user. 

\subsection{\textsc{Stackarator} operation}

\subsubsection{Stacking entire datacubes}
\label{wholecubestack}

By default \stack\ has two modes of operation. In the first mode, it can be used to stack the entire datacube, producing one single output spectrum.
\stack\ loops over each spaxel with a valid velocity in the input velocity field, resampling it onto a consistent grid using one-dimensional linear interpolation, and offsetting it by the velocity expected at the spaxel centroid position (from Eqn. \ref{velfield}). As such, if the input velocity field accurately predicts the velocity of the tracer then the output stacked spectrum will peak at a velocity of 0 \kms.

Given that (at least for galaxies, as used here) the velocity width of the emission line can be a substantial fraction of the total observed bandwidth, each velocity bin in the output may have included contributions from a different number of input spaxels, and thus will have a different signal-to-noise ratio. This is tracked by the code, and this information is available to the user. 

The final outputs of the code in this mode are thus the output velocities, the stacked spectrum sampled at these velocities, the noise expected, and the number of input spaxels which contributed to each velocity bin. We note that the noise estimate output by the code for each velocity bin has been propagated assuming each spaxel is independent, and the noise in the input datacube is gaussian. These assumptions are likely to be violated in typical interferemetric datacubes (where, for instance, the synthesised beam is typically oversampled by the pixel grid). As such in this work we recalculated the noise in the output spectrum, and thus how significant our line detections were, using a resampling technique as described in Section \ref{stack_describe}.

\subsubsection{Regional stacking}

The second mode of \stack\ allows it to stack specific regions of the input datacube, to for instance reveal the properties of weak lines as a function of radius within a galaxy.

In this mode a mask must be created, to define the stacking area. The code includes a helper function to create both radial elliptical masks, and wedges of a given angular size at user defined position angles. 

The \textsc{define\_region\_ellipse} function takes the RA and Dec position of the galaxy centre, along with the galaxy inclination and position angle. These are used to create elliptical boolean masks aligned with the galaxy disc, with a radial extent as specified by the user (see e.g. Figure \ref{fig:advert_radial_stack}). 

The \textsc{define\_region\_angrange} function takes the RA and Dec position of the galaxy centre, along with the position angle you wish to extract a region from, and an angular size for this wedge. Optionally one can include a radius below which the wedge will be truncated (e.g. to avoid the galaxy centre). These are used to create wedge-shaped boolean masks aligned with the supplied  position angle, with an angular extent as specified by the user. 

If more complex masks are required these can be created by the user as 2D boolean arrays, evaluated on the pixel grid of the datacube, and passed to the code manually. 

Once a mask has been created, stacking proceeds as described in Section \ref{wholecubestack} but within this mask only. Looping over different radii, for instance, would allow one to explore the variation of the stacked lines across the face of a galaxy (as demonstrated in Figure \ref{fig:advert_radial_stack}).

\section{\tco\ maps}

\begin{figure*}
\begin{subfigure}{0.32\textwidth}
	\includegraphics[width=\textwidth,trim=0cm 0cm 0inch 0cm,clip]{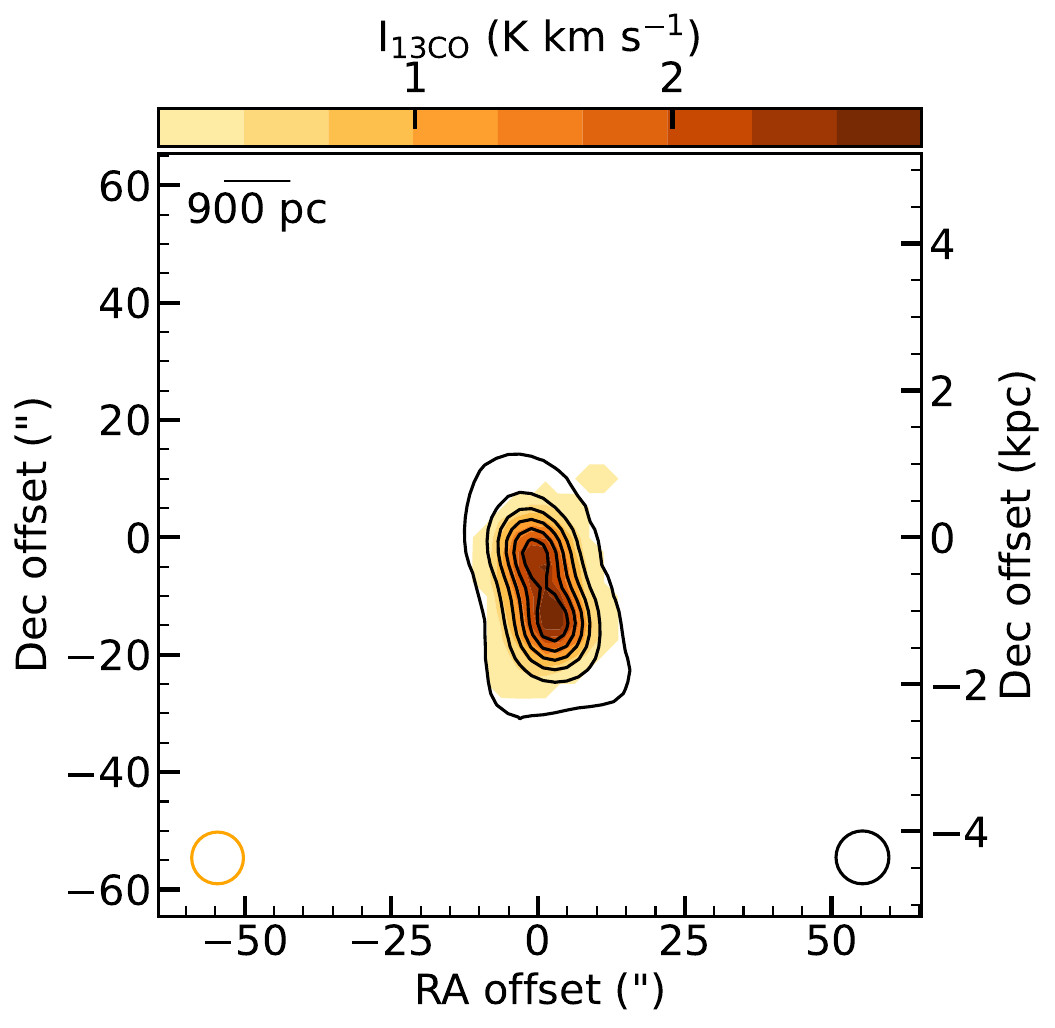}
    \caption{NGC4064}
\end{subfigure} 
\begin{subfigure}{0.32\textwidth}
	\includegraphics[width=\textwidth,trim=0cm 0cm 0inch 0cm,clip]{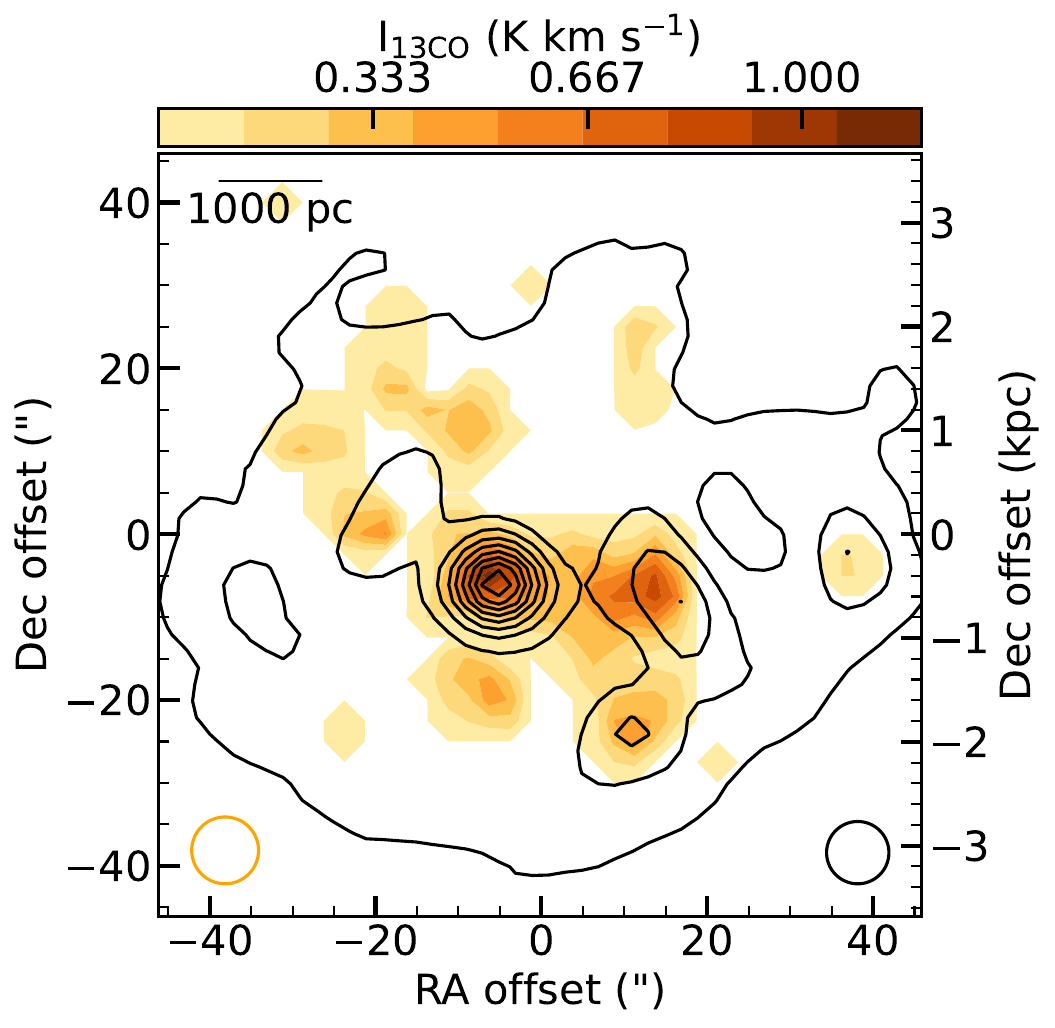}
    \caption{NGC4189}
\end{subfigure} 
\begin{subfigure}{0.32\textwidth}
	\includegraphics[width=\textwidth,trim=0cm 0cm 0inch 0cm,clip]{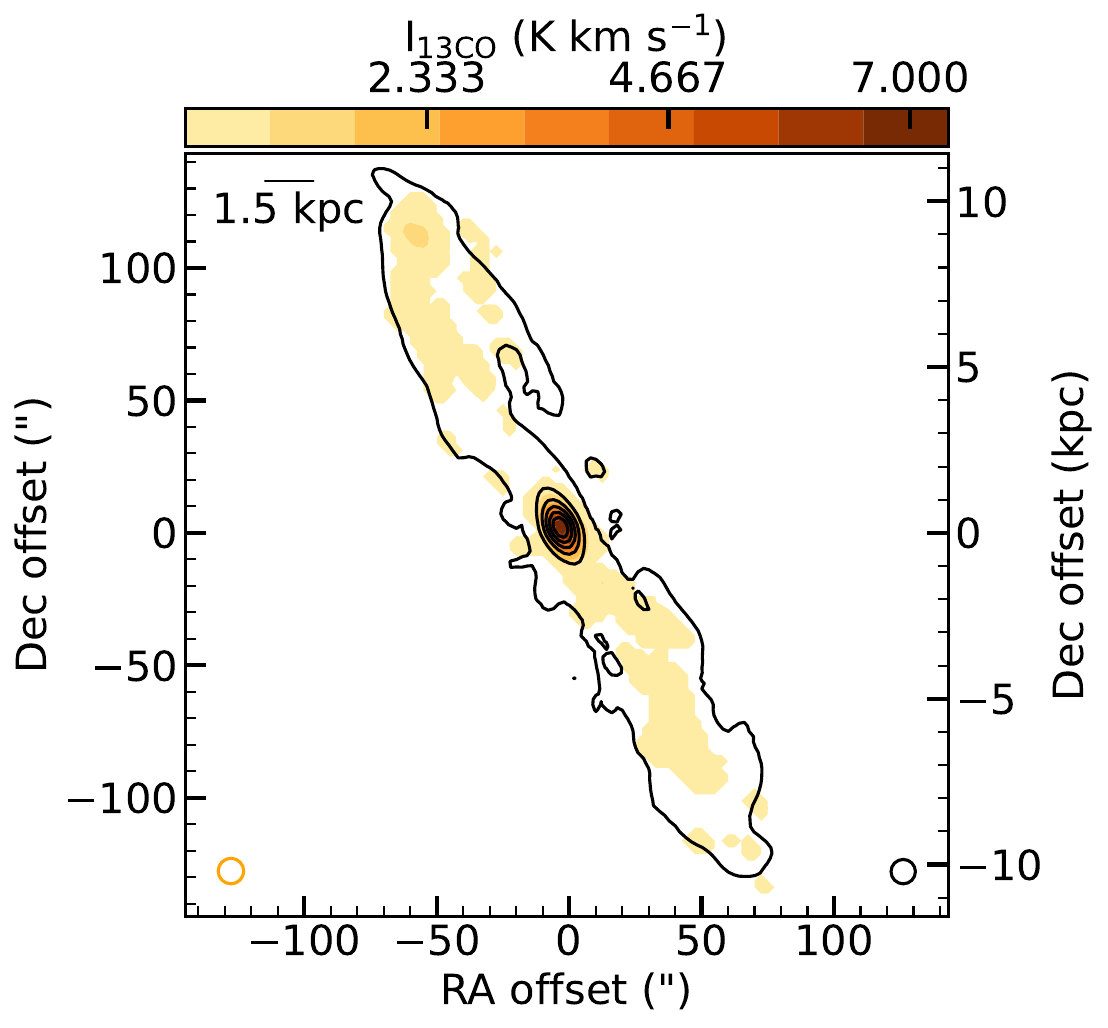}
    \caption{NGC4192}
\end{subfigure} 
\begin{subfigure}{0.32\textwidth}
	\includegraphics[width=\textwidth,trim=0cm 0cm 0inch 0cm,clip]{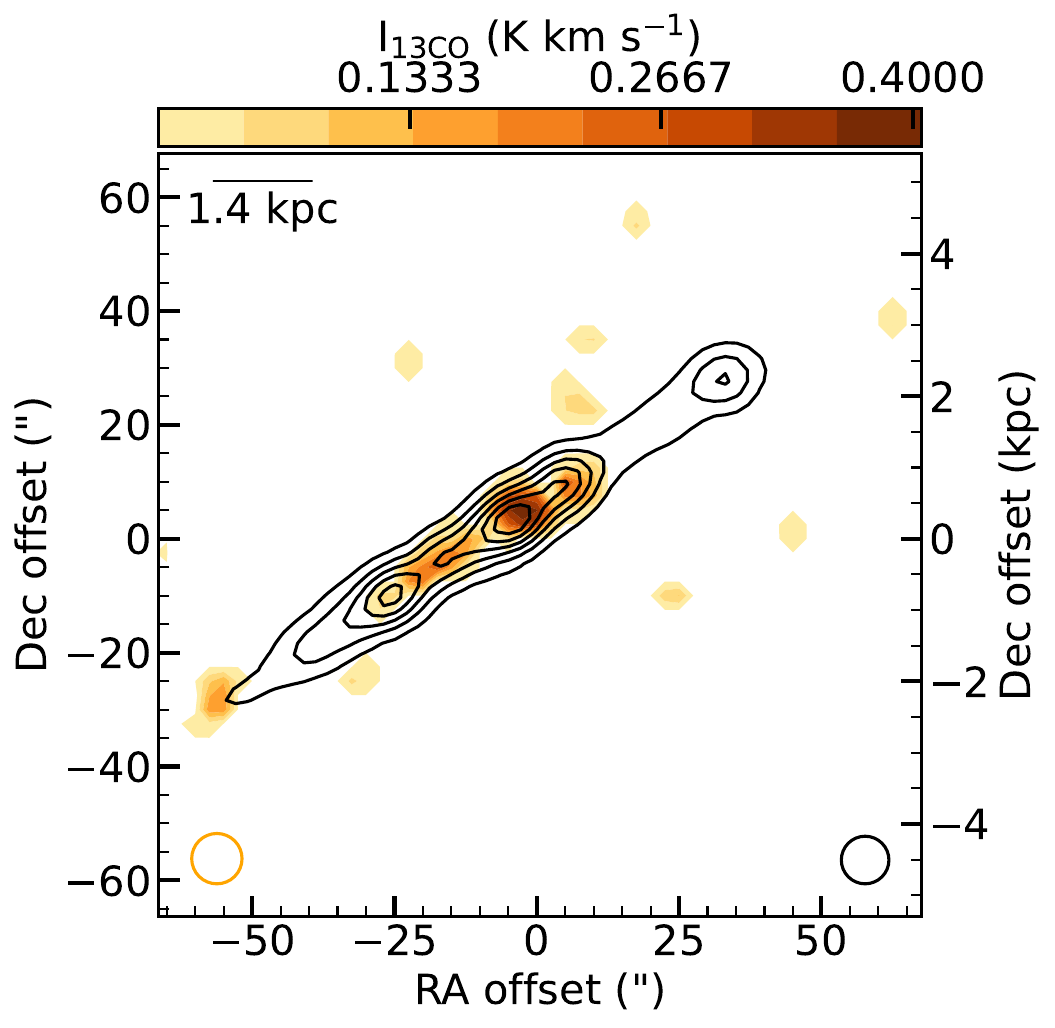}
    \caption{NGC4222}
\end{subfigure} 
\begin{subfigure}{0.32\textwidth}
	\includegraphics[width=\textwidth,trim=0cm 0cm 0inch 0cm,clip]{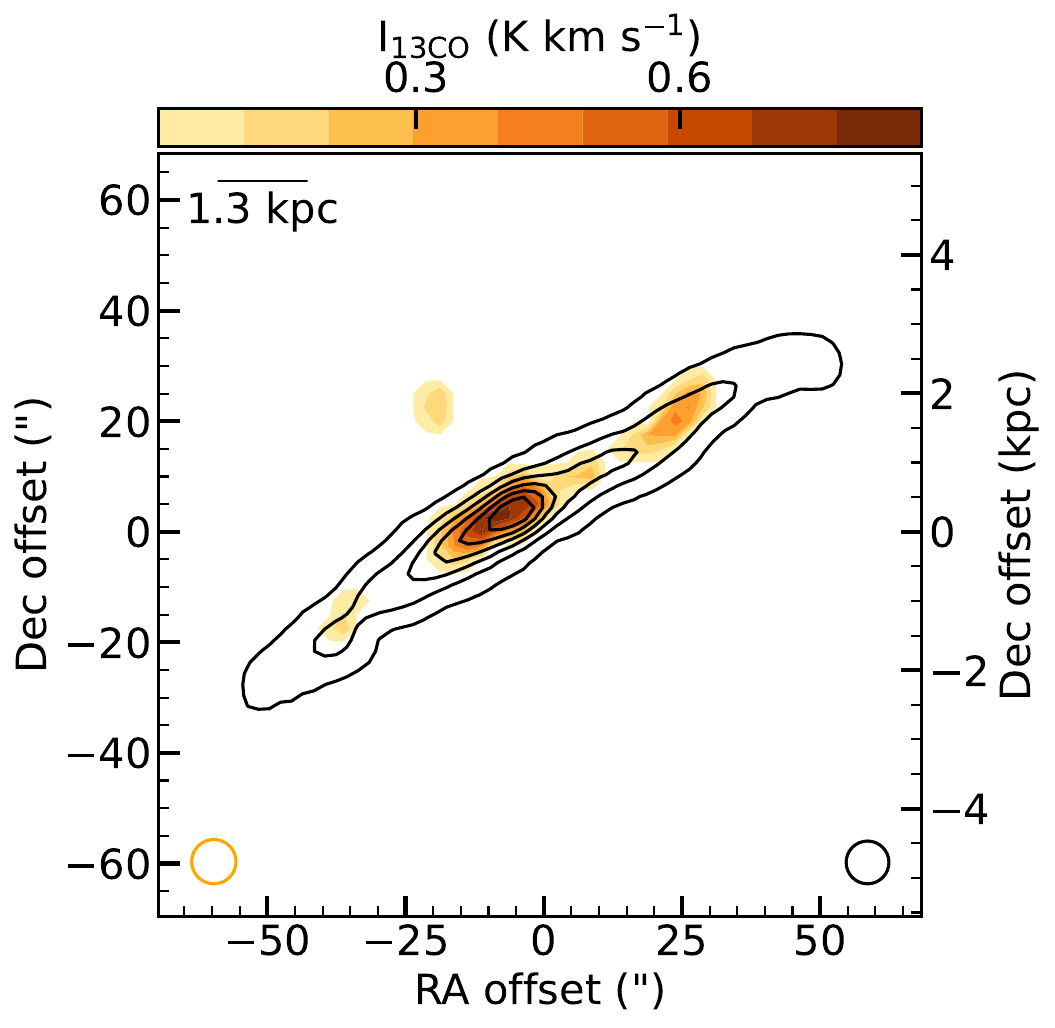}
    \caption{NGC4330}
\end{subfigure} 
\begin{subfigure}{0.32\textwidth}
	\includegraphics[width=\textwidth,trim=0cm 0cm 0inch 0cm,clip]{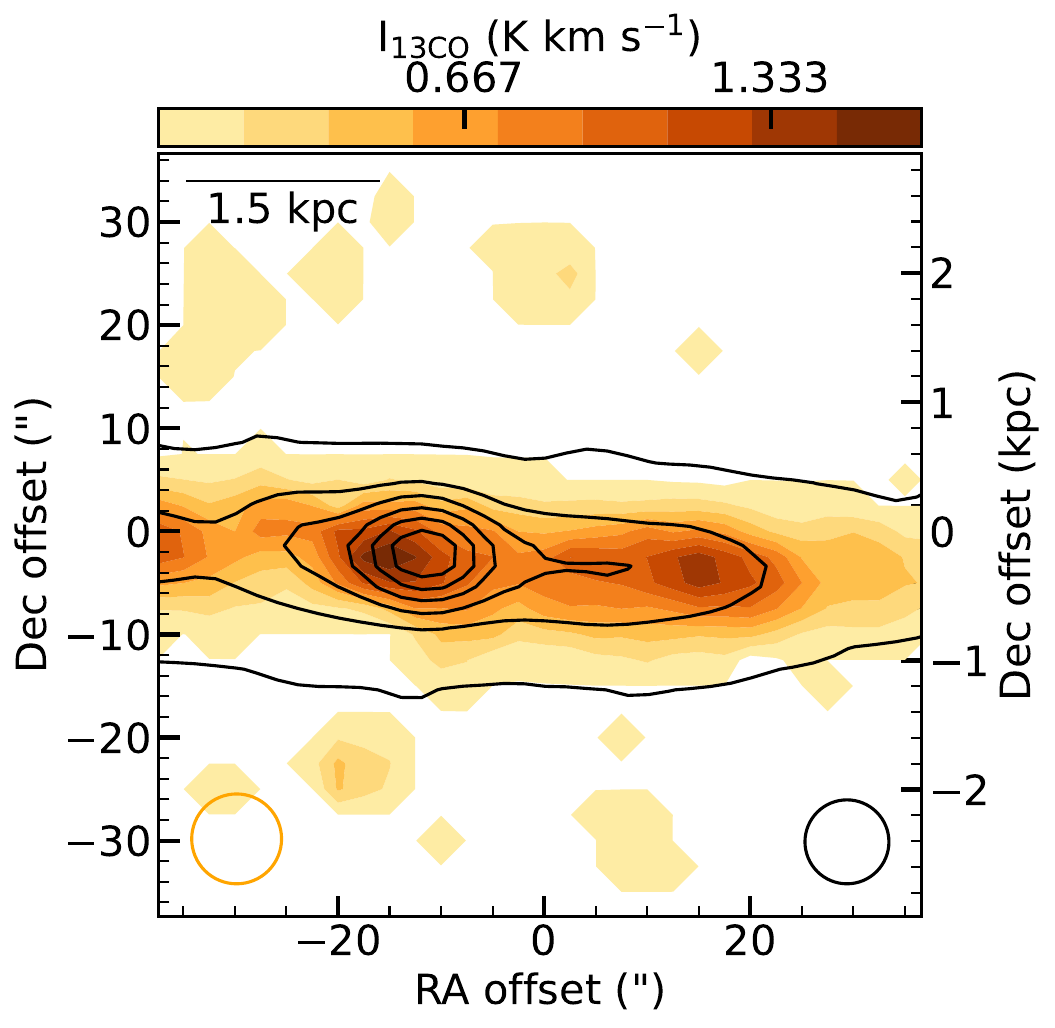}
    \caption{NGC4388}
\end{subfigure} 
\begin{subfigure}{0.32\textwidth}
	\includegraphics[width=\textwidth,trim=0cm 0cm 0inch 0cm,clip]{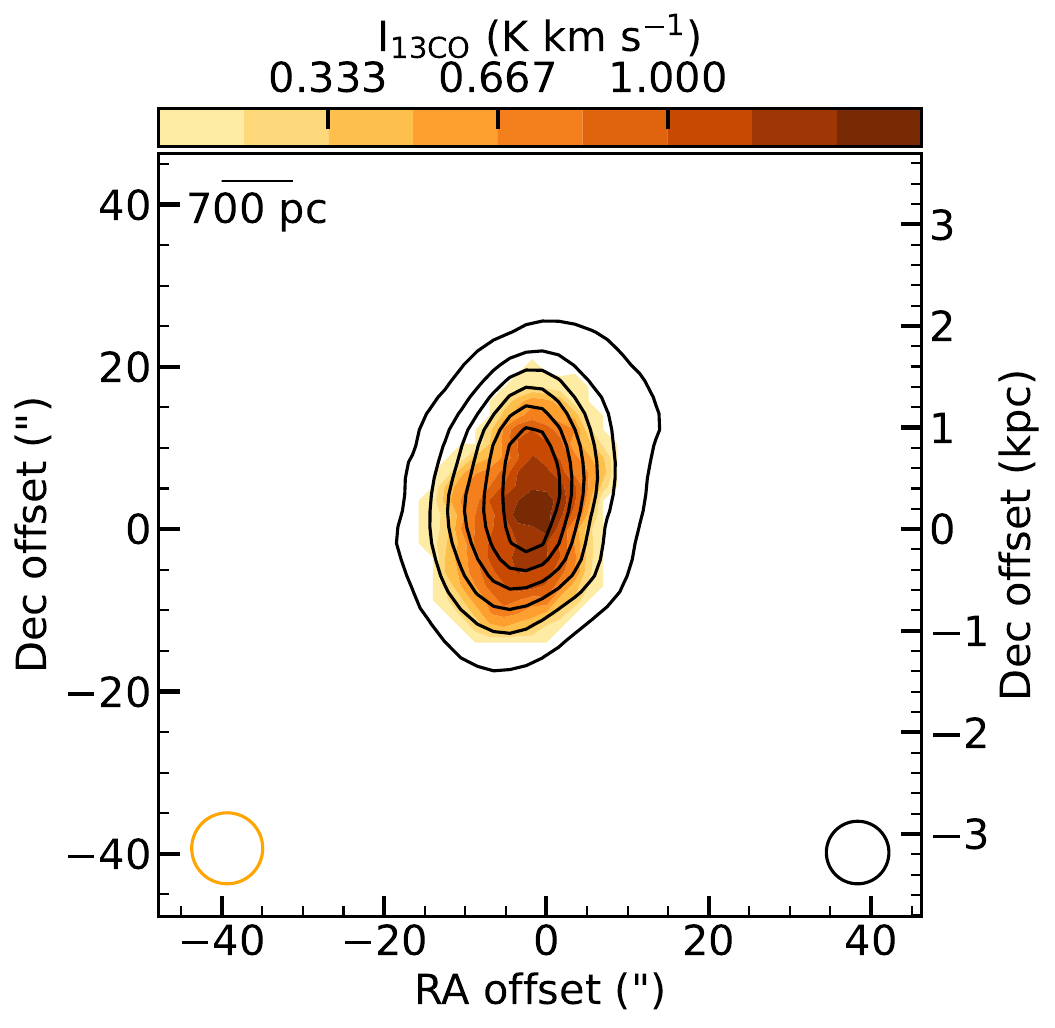}
    \caption{NGC4405}
\end{subfigure} 
\begin{subfigure}{0.32\textwidth}
	\includegraphics[width=\textwidth,trim=0cm 0cm 0inch 0cm,clip]{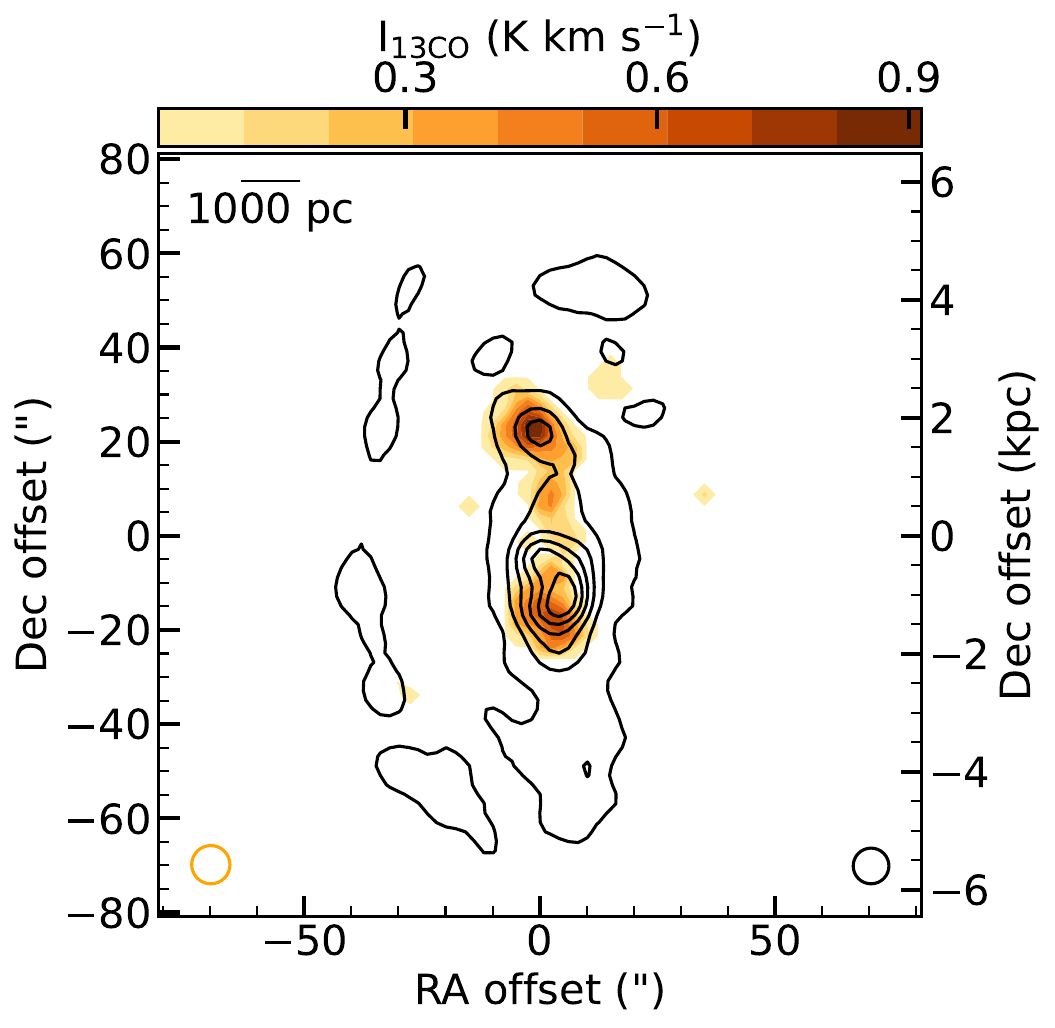}
    \caption{NGC4450}
\end{subfigure} 
\begin{subfigure}{0.32\textwidth}
	\includegraphics[width=\textwidth,trim=0cm 0cm 0inch 0cm,clip]{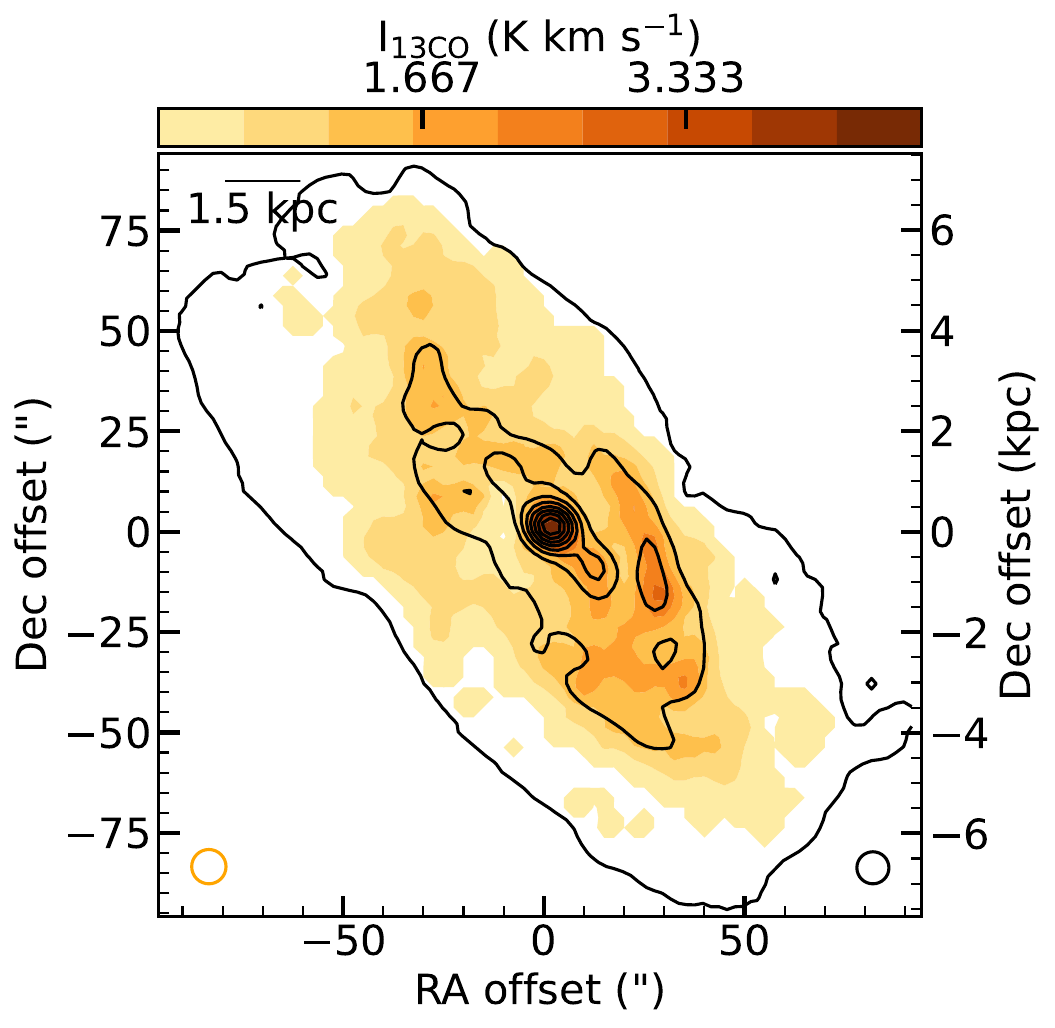}
	\label{tco_n4501}
    \caption{NGC4501}
\end{subfigure} 
\caption{\tco(2-1) integrated intensity (moment zero) maps for the systems where this line was detected without stacking. Note that NGC4302 and IC3392 are shown in Figure \ref{fig:directdetect}. In each case five black contours delineate the \co(2-1) emitting region, with contours starting at a column density of 5 \msun\,pc$^{-2}$, and spaced equally to the maximum.  The isotopologue emission is shown in orange in ten contours spaced from 10\% of the peak emission to 100\%. The corresponding  \co\ and isotopologue telescope beams are shown as black and orange ellipses in the bottom right and left corner of each image, respectively. }
\label{fig:directdetect_appendix}
\end{figure*}

\begin{figure*}
\begin{subfigure}{0.32\textwidth}
	\includegraphics[width=\textwidth,trim=0cm 0cm 0inch 0cm,clip]{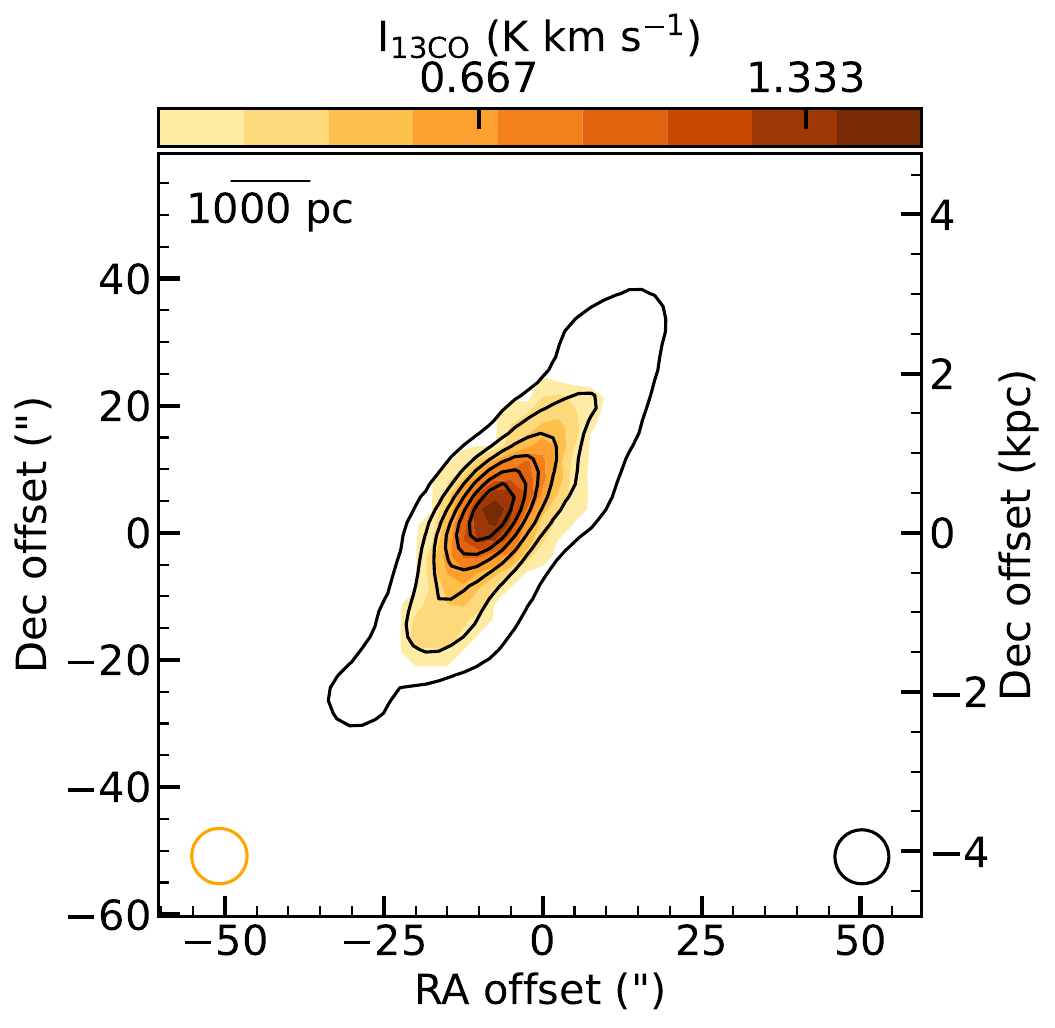}
    \caption{NGC4522}
\end{subfigure} 
\begin{subfigure}{0.32\textwidth}
	\includegraphics[width=\textwidth,trim=0cm 0cm 0inch 0cm,clip]{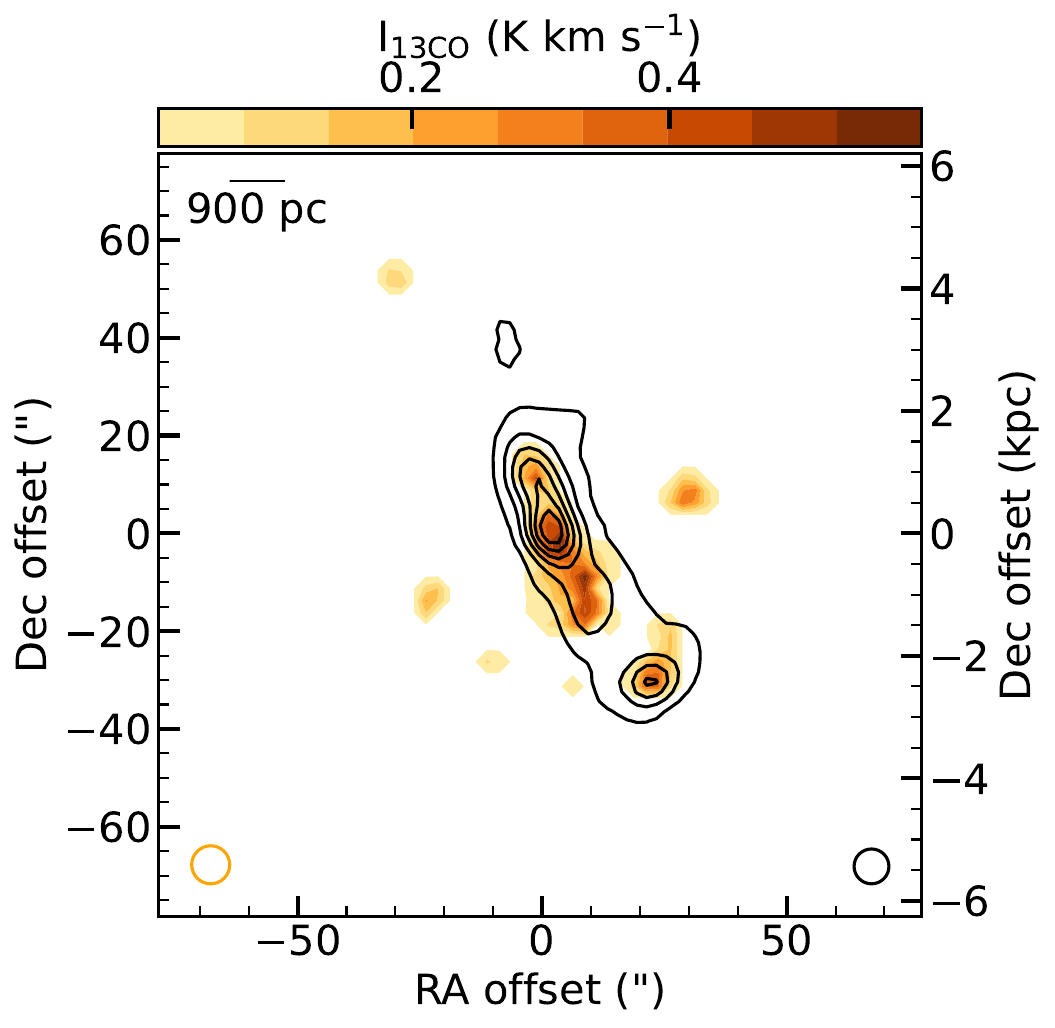}
    \caption{NGC4532}
\end{subfigure} 
\begin{subfigure}{0.32\textwidth}
	\includegraphics[width=\textwidth,trim=0cm 0cm 0inch 0cm,clip]{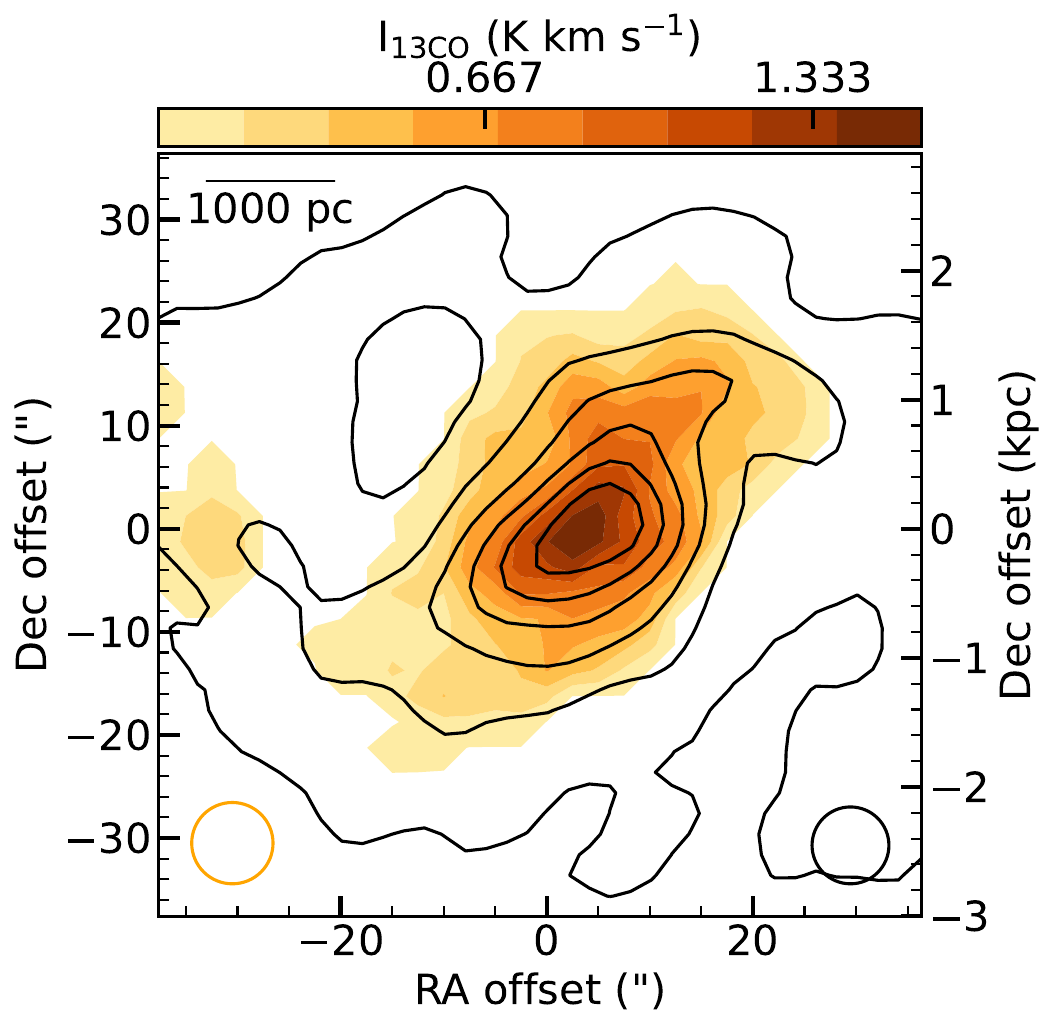}
    \caption{NGC4567}
\end{subfigure} 
\begin{subfigure}{0.32\textwidth}
	\includegraphics[width=\textwidth,trim=0cm 0cm 0inch 0cm,clip]{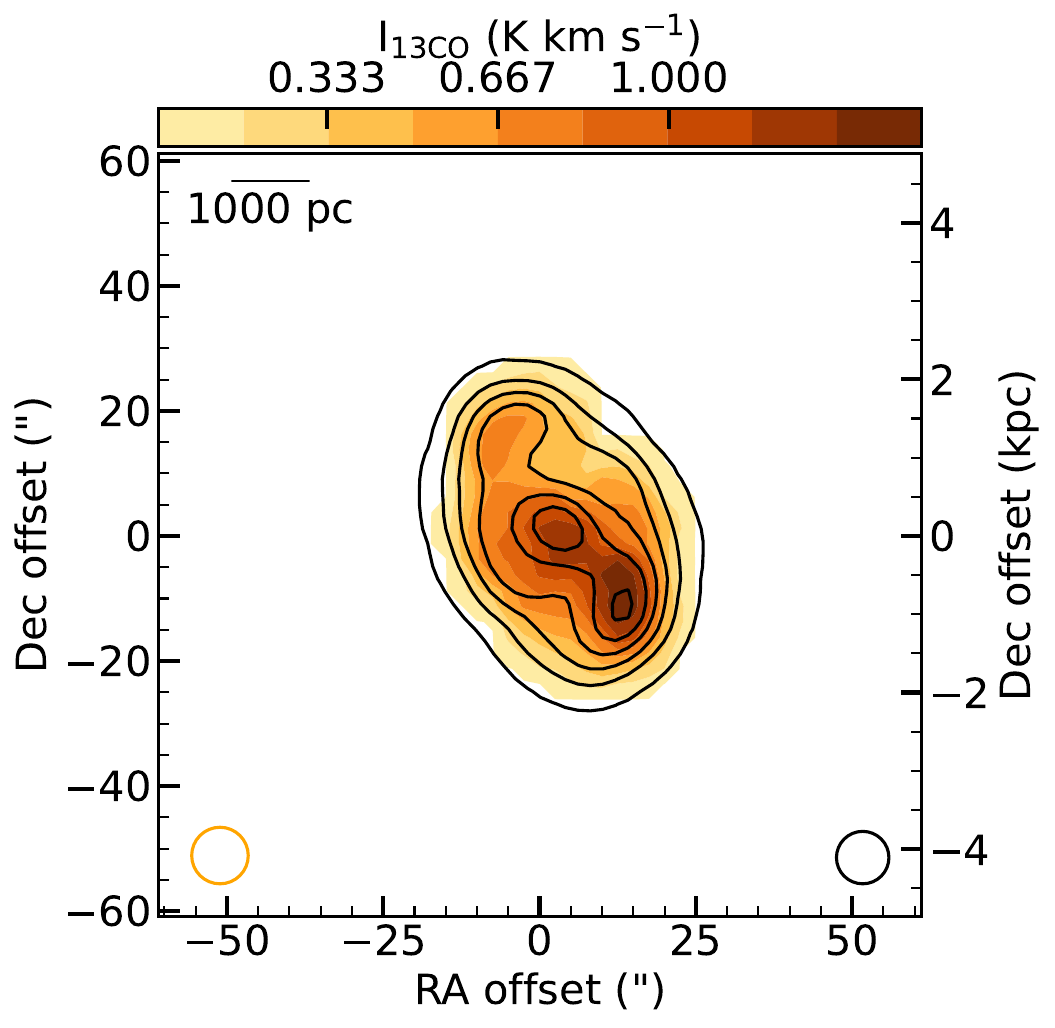}
    \caption{NGC4580}
\end{subfigure} 
\begin{subfigure}{0.32\textwidth}
	\includegraphics[width=\textwidth,trim=0cm 0cm 0inch 0cm,clip]{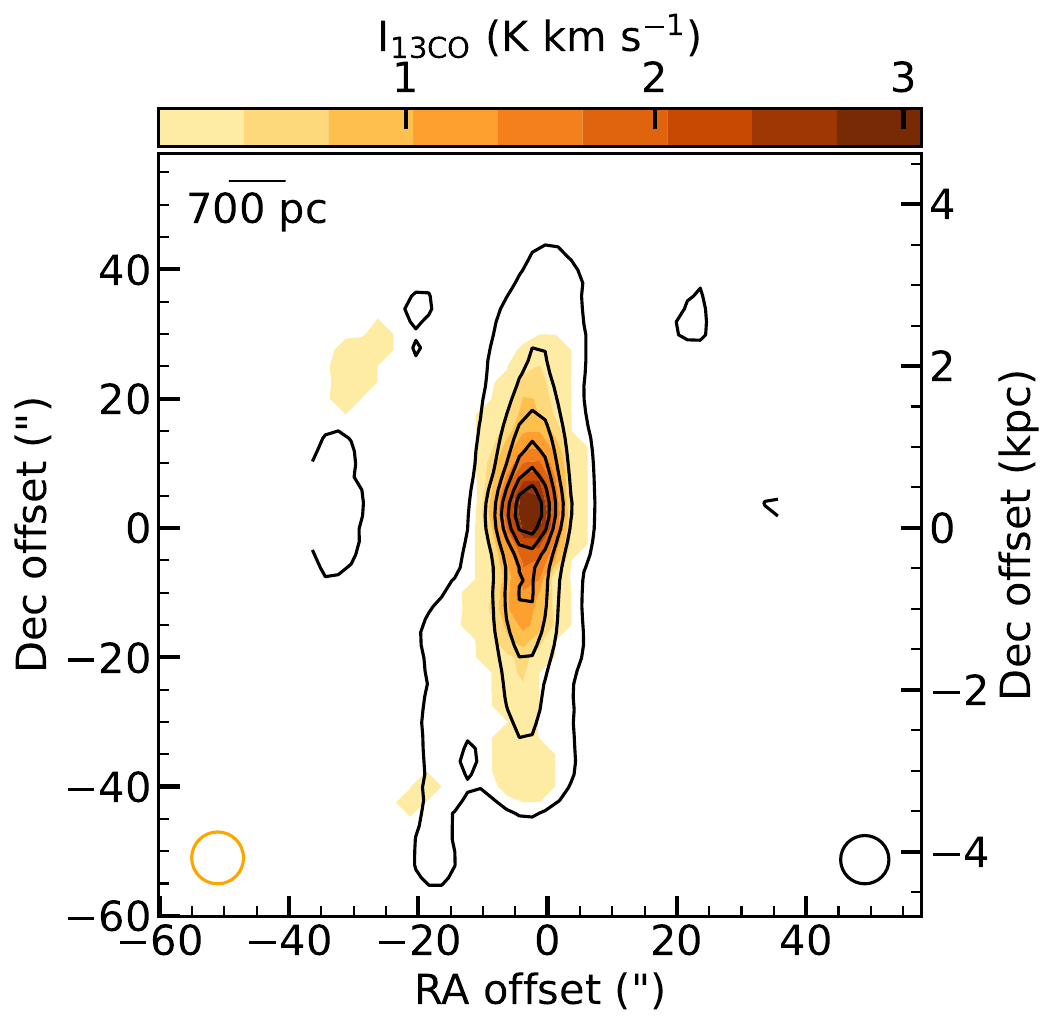}
    \caption{NGC4607}
\end{subfigure} 
\begin{subfigure}{0.32\textwidth}
	\includegraphics[width=\textwidth,trim=0cm 0cm 0inch 0cm,clip]{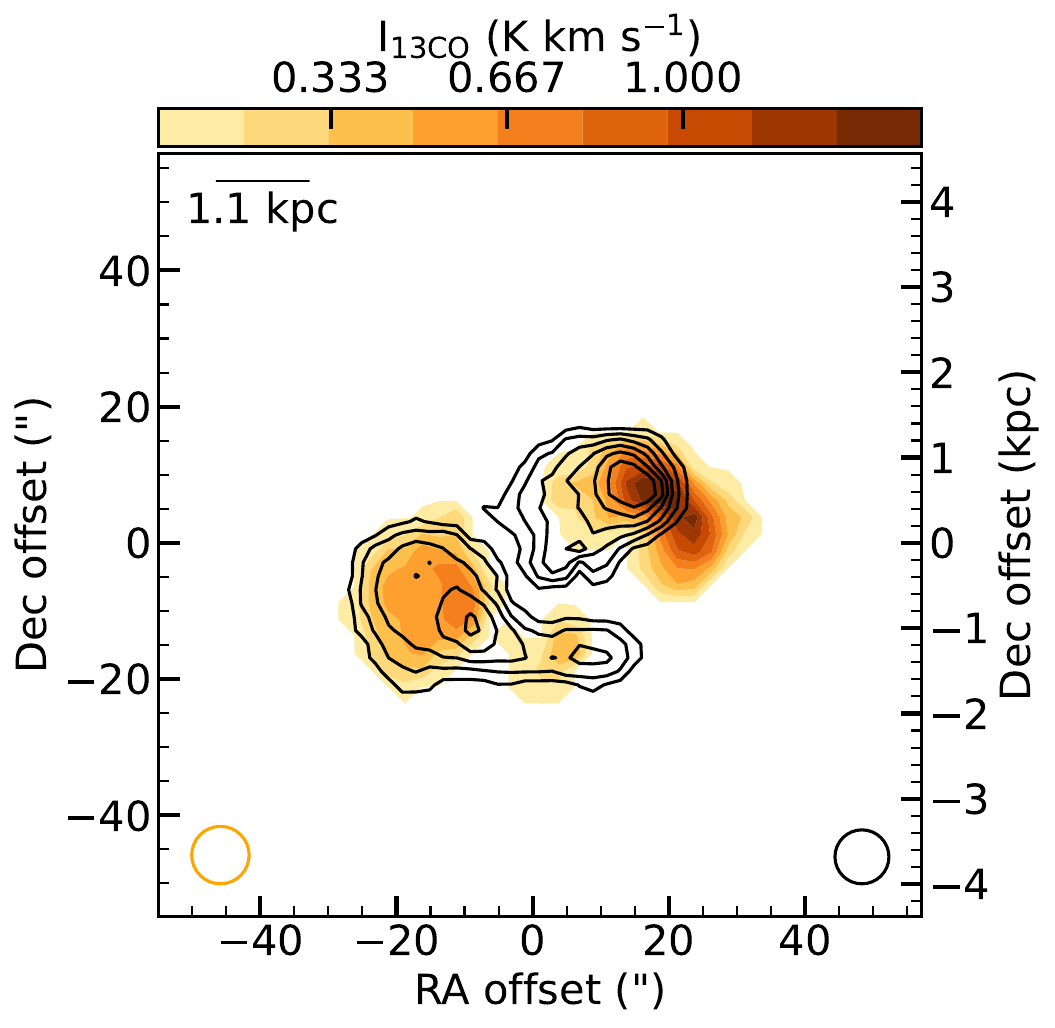}
    \caption{NGC4651}
\end{subfigure} 
\begin{subfigure}{0.32\textwidth}
	\includegraphics[width=\textwidth,trim=0cm 0cm 0inch 0cm,clip]{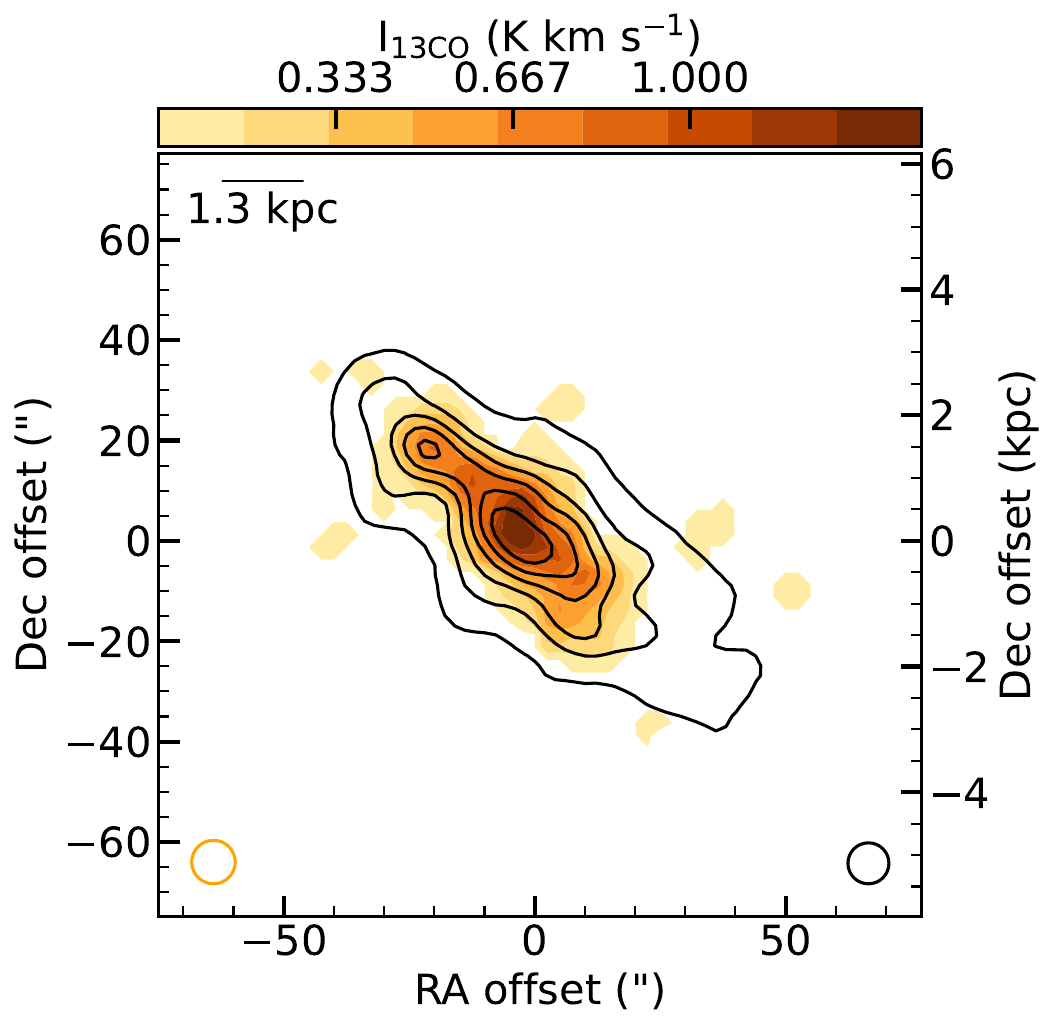}
    \caption{NGC4808}
\end{subfigure} 
\caption{As \protect Figure \ref{fig:directdetect_appendix}. Note that NGC4568 is shown in Figure \ref{fig:directdetect} }
\label{fig:directdetect_appendix2}
\end{figure*}

\section{Stacked line fluxes}
\begin{table*}
\caption{CO isotopologue fluxes measured from entire galaxy stacks}
\begin{tabular}{lrlllrlllrll}
\hline
Galaxy & \, & $\int S_{\nu} \delta V_{\mathrm{CO}}$ & $\Delta\int S_{\nu} \delta V_{\mathrm{CO}}$ & N$_{\rm g,CO}$ & \vspace{0pt} & $\int S_{\nu} \delta V_{\mathrm{13CO}}$ & $\Delta\int S_{\nu} \delta V_{\mathrm{13CO}}$ & N$_{\rm g,13CO}$ & \hspace{0pt} & $\int S_{\nu} \delta V_{\mathrm{C18O}}$ & $\Delta\int S_{\nu} \delta V_{\mathrm{C18O}}$ \\
 &  & (Jy km s$^{-1}$) & (Jy km s$^{-1}$) &  &  & (Jy km s$^{-1}$) & (Jy km s$^{-1}$) &  &  & (Jy km s$^{-1}$) & (Jy km s$^{-1}$) \\
(1) && (2) &(3) &(4) &&(5) &(6) &(7)&&(8) &(9)\\ \hline
IC3392 &  & 273.4 & 0.9 & 2 &  & 19.4 & 1.4 & 1 &  & 4.6 & 0.6 \\
NGC4064 &  & 299.5 & 0.4 & 1 &  & 22.6 & 0.3 & 1 & < & 3.4 & -- \\
NGC4189 &  & 485.1 & 3.6 & 2 &  & 36.0 & 2.8 & 1 & < & 5.0 & -- \\
NGC4192 &  & 1863.9 & 3.2 & 2 &  & 108.4 & 10.8 & 2 & < & 7.4 & -- \\
NGC4216 &  & 962.0 & 4.7 & 2 &  & 67.1 & 3.8 & 1 & < & 6.9 & -- \\
NGC4222 &  & 126.2 & 0.5 & 1 &  & 3.9 & 0.4 & 1 & < & 0.9 & -- \\
NGC4254 &  & 7858.5 & 6.1 & 2 &  & -- & -- & 1 &  & 24.3 & 2.7 \\
NGC4293 &  & 785.2 & 2.0 & 2 &  & -- & -- & 1 & < & 2.8 & -- \\
NGC4294 &  & 62.2 & 0.7 & 1 &  & 2.7 & 0.4 & 1 & < & 1.2 & -- \\
NGC4298 &  & 1267.2 & 3.2 & 2 &  & -- & -- & 1 &  & 3.5 & 0.6 \\
NGC4299 &  & 39.5 & 1.6 & 1 & < & 1.7 & -- & -- & < & 1.4 & -- \\
NGC4302 &  & 1494.2 & 3.0 & 2 &  & 125.8 & 3.7 & 2 & < & 11.8 & -- \\
NGC4321 &  & 7298.1 & 3.0 & 2 &  & -- & -- & 1 & < & 13.4 & -- \\
NGC4330 &  & 238.6 & 0.5 & 2 &  & 9.5 & 0.5 & 1 & < & 1.4 & -- \\
NGC4351 &  & 68.3 & 2.2 & 1 &  & 1.5 & 0.3 & 1 & < & 1.7 & -- \\
NGC4380 &  & 403.9 & 2.7 & 2 &  & 24.0 & 1.8 & 1 &  & 7.2 & 1.2 \\
NGC4383 &  & 245.2 & 2.2 & 2 & < & 3.6 & -- & -- & < & 3.1 & -- \\
NGC4388 &  & 829.4 & 0.3 & 2 &  & 23.3 & 0.6 & 1 & < & 2.2 & -- \\
NGC4394 &  & 92.6 & 2.6 & 2 &  & 3.4 & 0.6 & 1 & < & 1.3 & -- \\
NGC4396 &  & 105.5 & 1.6 & 2 &  & 10.1 & 0.5 & 2 & < & 1.7 & -- \\
NGC4405 &  & 216.2 & 0.7 & 1 &  & 16.0 & 0.5 & 1 & < & 1.9 & -- \\
NGC4419 &  & 1415.9 & 1.2 & 2 &  & 9.6 & 0.4 & 1 &  & 5.6 & 0.8 \\
NGC4424 &  & 224.6 & 1.4 & 1 &  & -- & -- & 1 & < & 3.3 & -- \\
NGC4450 &  & 458.5 & 2.5 & 2 &  & 27.6 & 1.6 & 1 & < & 4.6 & -- \\
NGC4457 &  & 1080.6 & 2.0 & 2 &  & -- & -- & 1 & < & 11.8 & -- \\
NGC4501 &  & 5233.6 & 10.6 & 2 &  & 420.3 & 5.8 & 2 &  & 13.6 & 2.3 \\
NGC4522 &  & 231.0 & 0.8 & 2 &  & 14.8 & 0.8 & 1 & < & 1.9 & -- \\
NGC4532 &  & 165.1 & 3.0 & 2 &  & 7.8 & 0.7 & 1 & < & 3.3 & -- \\
NGC4533 &  & 15.5 & 0.3 & 1 & < & 0.6 & -- & -- & < & 0.4 & -- \\
NGC4535 &  & 2624.9 & 5.8 & 2 &  & -- & -- & 1 & < & 13.7 & -- \\
NGC4536 &  & 2375.2 & 11.7 & 2 &  & -- & -- & 1 & < & 3.9 & -- \\
NGC4548 &  & 861.4 & 11.9 & 2 &  & -- & -- & 1 & < & 3.4 & -- \\
NGC4561 &  & 13.1 & 1.0 & 1 & < & 1.3 & -- & -- & < & 0.9 & -- \\
NGC4567 &  & 743.4 & 1.4 & 2 &  & 28.7 & 0.9 & 1 & < & 2.5 & -- \\
NGC4568 &  & 2772.2 & 6.4 & 2 &  & 251.8 & 4.4 & 2 &  & 37.1 & 3.5 \\
NGC4569 &  & 3945.4 & 3.5 & 2 &  & -- & -- & 1 & < & 15.0 & -- \\
NGC4579 &  & 2087.7 & 7.1 & 2 &  & -- & -- & 1 & < & 3.5 & -- \\
NGC4580 &  & 393.2 & 2.7 & 2 &  & 34.6 & 1.2 & 2 & < & 2.5 & -- \\
NGC4606 &  & 143.9 & 1.0 & 1 &  & 9.5 & 0.3 & 1 & < & 1.5 & -- \\
NGC4607 &  & 440.6 & 1.8 & 2 &  & 29.5 & 0.9 & 1 & < & 4.1 & -- \\
NGC4651 &  & 578.2 & 3.5 & 2 &  & 40.2 & 1.7 & 1 & < & 4.2 & -- \\
NGC4654 &  & 2187.5 & 9.6 & 2 &  & -- & -- & 1 &  & 8.8 & 1.1 \\
NGC4689 &  & 1088.7 & 3.5 & 2 &  & -- & -- & 1 & < & 3.6 & -- \\
NGC4694 &  & 113.4 & 0.4 & 2 &  & -- & -- & 1 & < & 1.5 & -- \\
NGC4698 &  & 51.6 & 1.5 & 1 & < & 1.9 & -- & -- & < & 1.6 & -- \\
NGC4713 &  & 226.9 & 2.6 & 1 &  & 13.0 & 2.0 & 1 & < & 3.5 & -- \\
NGC4772 &  & 29.2 & 0.6 & 1 & < & 0.5 & -- & -- & < & 0.4 & -- \\
NGC4808 &  & 571.9 & 1.5 & 2 &  & 36.3 & 1.4 & 1 & < & 3.5 & -- \\
\hline
\end{tabular}
\\ \textit{Notes:} Column 1 indicates the galaxy, while Columns 2 and 3 indicate the \co\ integrated intensity and its uncertainty based on stacks of the entire galaxy emission. Upper limits are quoted at the 3$\sigma$ level assuming the isotopologue has the same velocity width as the \co\ line, and indicated with a $<$ symbol. Column 4 indicates how many gaussian components were used to fit the stacked line. Columns 5+6+7 and 8+9 contain the same information, but for \tco\ and \ceo\ respectively. Single gaussian fits were used for all \ceo\ lines. Objects without \tco\ observations are indicated by dashes in Column 5. \label{Table1}
\end{table*}

\subsection{Line ratios}
\begin{table*}
\caption{Global integrated CO isotopologue ratios for our galaxies.}
\begin{tabular}{lrllrllrll}
\hline
Galaxy & \, & \co/\tco & $\Delta$\co/\tco & \vspace{0pt} & \co/\ceo & $\Delta$\co/\ceo & \hspace{0pt} & \tco/\ceo & $\Delta$\tco/\ceo \\
(1) && (2) &(3) &&(4) &(5) &&(6) &(7)\\ \hline
IC3392 &  & 12.9 & 0.9 &  & 60 & 8 &  & 4 & 1 \\
NGC4064 &  & 12.1 & 0.2 & > & 88 & -- & > & 7 & -- \\
NGC4189 &  & 12.3 & 1.0 & > & 97 & -- & > & 7 & -- \\
NGC4192 &  & 15.7 & 1.6 & > & 251 & -- & > & 15 & -- \\
NGC4216 &  & 13.1 & 0.8 & > & 139 & -- & > & 10 & -- \\
NGC4222 &  & 29.8 & 2.7 & > & 135 & -- & > & 4 & -- \\
NGC4254 &  & -- & -- &  & 323 & 36 &  & -- & -- \\
NGC4293 &  & -- & -- & > & 281 & -- &  & -- & -- \\
NGC4294 &  & 20.7 & 3.1 & > & 53 & -- & > & 2 & -- \\
NGC4298 &  & -- & -- &  & 358 & 65 &  & -- & -- \\
NGC4299 & > & 21.3 & -- & > & 28 & -- & > & 1 & -- \\
NGC4302 &  & 10.9 & 0.3 & > & 127 & -- & > & 11 & -- \\
NGC4321 &  & -- & -- & > & 543 & -- &  & -- & -- \\
NGC4330 &  & 22.9 & 1.3 & > & 173 & -- & > & 7 & -- \\
NGC4351 &  & 42.7 & 8.6 & > & 41 & -- & > & 1 & -- \\
NGC4380 &  & 15.4 & 1.2 &  & 56 & 9 &  & 3 & 1 \\
NGC4383 & > & 62.1 & -- & > & 80 & -- & > & 2 & -- \\
NGC4388 &  & 32.5 & 0.9 & > & 371 & -- & > & 10 & -- \\
NGC4394 &  & 25.2 & 4.6 & > & 71 & -- & > & 3 & -- \\
NGC4396 &  & 9.6 & 0.5 & > & 61 & -- & > & 6 & -- \\
NGC4405 &  & 12.4 & 0.4 & > & 113 & -- & > & 8 & -- \\
NGC4419 &  & -- & 6.0 &  & 251 & 35 &  & -- & 0 \\
NGC4424 &  & -- & -- & > & 68 & -- &  & -- & -- \\
NGC4450 &  & 15.2 & 0.9 & > & 100 & -- & > & 6 & -- \\
NGC4457 &  & -- & -- & > & 92 & -- &  & -- & -- \\
NGC4501 &  & 11.4 & 0.2 &  & 383 & 64 &  & 31 & 5 \\
NGC4522 &  & 14.3 & 0.7 & > & 120 & -- & > & 8 & -- \\
NGC4532 &  & 19.3 & 1.8 & > & 50 & -- & > & 2 & -- \\
NGC4533 & > & 24.0 & -- & > & 39 & -- & > & 2 & -- \\
NGC4535 &  & -- & -- & > & 191 & -- &  & -- & -- \\
NGC4536 &  & -- & -- & > & 604 & -- &  & -- & -- \\
NGC4548 &  & -- & -- & > & 251 & -- &  & -- & -- \\
NGC4561 & > & 8.9 & -- & > & 15 & -- & > & 1 & -- \\
NGC4567 &  & 23.6 & 0.7 & > & 301 & -- & > & 12 & -- \\
NGC4568 &  & 10.1 & 0.2 &  & 75 & 7 &  & 7 & 1 \\
NGC4569 &  & -- & -- & > & 263 & -- &  & -- & -- \\
NGC4579 &  & -- & -- & > & 602 & -- &  & -- & -- \\
NGC4580 &  & 10.4 & 0.4 & > & 157 & -- & > & 14 & -- \\
NGC4606 &  & 13.9 & 0.5 & > & 94 & -- & > & 6 & -- \\
NGC4607 &  & 13.7 & 0.4 & > & 109 & -- & > & 7 & -- \\
NGC4651 &  & 13.1 & 0.6 & > & 136 & -- & > & 9 & -- \\
NGC4654 &  & -- & -- &  & 247 & 31 &  & -- & -- \\
NGC4689 &  & -- & -- & > & 299 & -- &  & -- & -- \\
NGC4694 &  & -- & -- & > & 76 & -- &  & -- & -- \\
NGC4698 & > & 24.5 & -- & > & 33 & -- & > & 2 & -- \\
NGC4713 &  & 15.9 & 2.4 & > & 65 & -- & > & 4 & -- \\
NGC4772 & > & 56.5 & -- & > & 65 & -- & > & 1 & -- \\
NGC4808 &  & 14.4 & 0.5 & > & 163 & -- & > & 10 & -- \\
\hline
\end{tabular}
\\ \textit{Notes:} Column 1 indicates the galaxy, while Columns 2 and 3, 4 and 5 and 6 and 7 indicate the value of the indicated ratio, and its uncertainty respectively. Ratios are calculated from line intensities measured in units of K\,km\,s$^{-1}$. Where limits on ratios are indicated, these are at the 3$\sigma$ level. \label{Table2}
\end{table*}

\section{\co/\ceo\ radial profiles}
\begin{figure*}
	\includegraphics[width=1\textwidth,trim=0cm 0cm 0cm 0cm,clip]{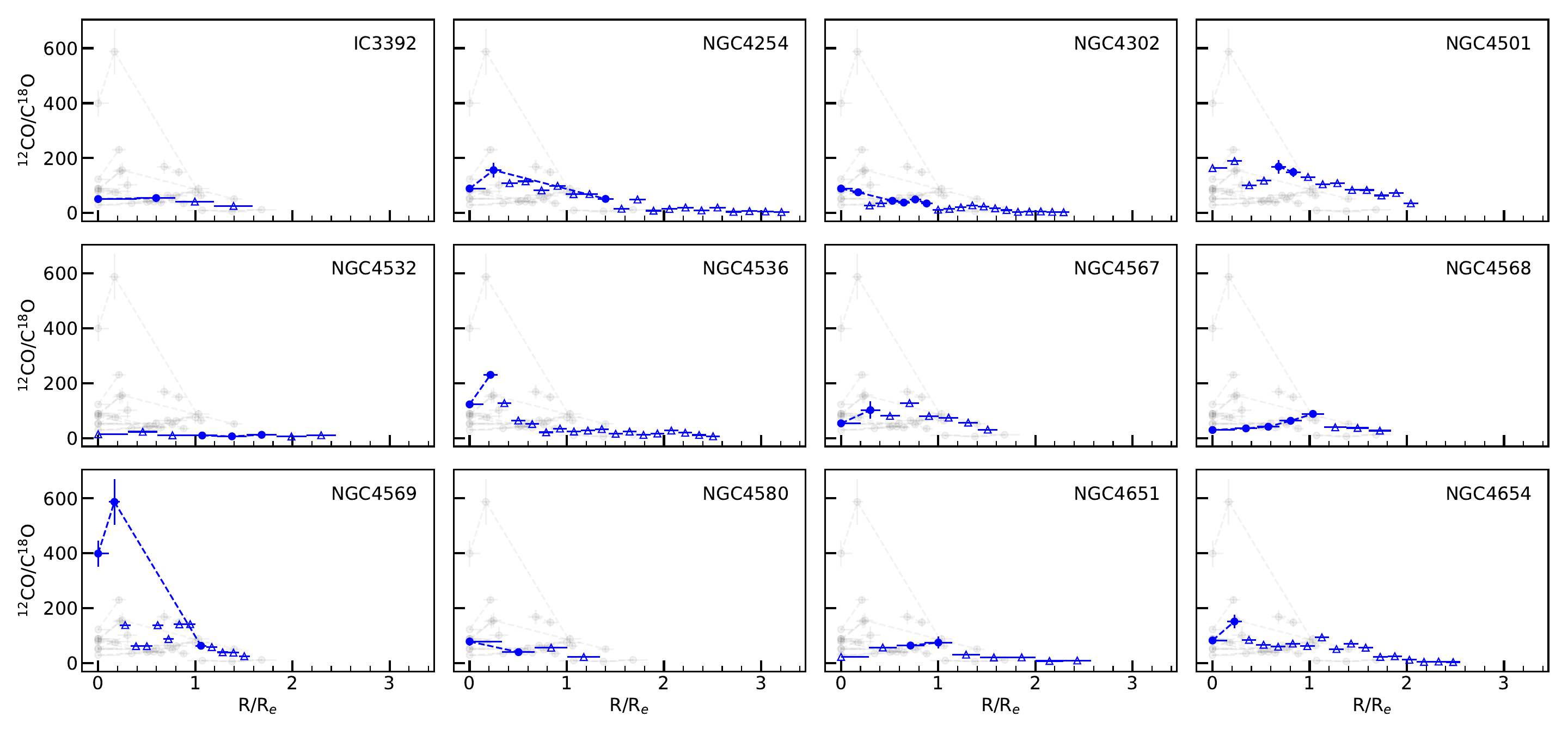}\\
	\caption{As Figure \ref{fig:radial_stack_r21}, but showing \co/\ceo\ line ratio radial profiles.}
	\label{fig:radial_stack_rco_ceo}
\end{figure*}

\end{document}